\documentclass[twocolumn,longbibliography]{revtex4-2}
\usepackage{graphicx} 
\usepackage{amsmath}
\usepackage{amsfonts}
\usepackage{braket}
\usepackage{bm}
\usepackage{tikz}
\usetikzlibrary{patterns,calc}
\usetikzlibrary{decorations.pathmorphing}
\usepackage{xcolor}
\usepackage{comment}
\usetikzlibrary{arrows.meta,patterns,patterns.meta}
\usepackage[dvipsnames]{xcolor}
\usepackage{hyperref}
\hypersetup{
    colorlinks,
    linkcolor={blue!50!black},
    citecolor={blue!50!black},
    urlcolor={blue!80!black},
    pdfborderstyle={/S/U/W 0}
}

\begin{document}
\title{Wavefront Curvature and Transverse Atomic Motion in Time-Resolved Atom Interferometry: Impact and Mitigation}
\author{Noam Mouelle}
\email[corresponding author: ]{ndm33@cam.ac.uk}
\author{Jeremiah Mitchell}
\author{Valerie Gibson}
\author{Ulrich Schneider}

 \affiliation{Cavendish Laboratory, University of Cambridge, J.J. Thomson Avenue, Cambridge CB3 0US, UK}
\date{\today}

\begin{abstract}
    Time-resolved atom interferometry, as employed in applications such as gravitational wave detection and searches for ultra-light dark matter, requires precise control over systematic effects. In this work, we investigate phase 
    noise arising from shot-to-shot fluctuations in the atoms' transverse motion in the presence of the wavefront 
    curvature of the interferometer beam, and analyse its dependence on the laser-beam geometry in long-baseline, large-momentum-transfer atom interferometers. We use a 
    semi-classical framework to derive analytical expressions for the effective phase perturbation in position-averaged measurements and 
    validate them using Monte Carlo simulations.
    Applied to 100-m and 1-km atom gradiometers representative of next-generation experiments, the model shows that configurations maximizing pulse efficiency also amplify curvature-induced phase noise, requiring micron-level control of the atom cloud's centre-of-mass position and sub-micron-per-second control of its centre-of-mass velocity to achieve sub-$10^{-5}$ rad phase stability. 
Alternative beam geometries can suppress this noise by up to two orders of magnitude, but at the cost of reduced pulse efficiency.
To address this limitation, we propose a mitigation strategy based on position-resolved phase-shift readout, which empirically learns and corrects the wavefront-induced bias from measurable quantities such as the phase-shift gradient and final cloud position.
This approach restores high-sensitivity operation in the maximum-pulse-efficiency configuration without detailed beam characterisation, providing a practical route towards next-generation, time-resolved atom interferometers operating at the $10^{-5}$ rad noise level.
\end{abstract}

\maketitle


\section{Introduction}

Light-pulse atom interferometry~\cite{YOUNG1997363, Borde1989} exploits interference between atomic wave packets travelling along spatially separated paths. Laser pulses coherently split, redirect, and recombine the wave packets, enabling precision measurements of external forces and fields. 
Atom interferometers have been successfully used in diverse applications, including inertial sensing~\cite{peters1999, PhysRevLett.78.2046, PhysRevA.65.033608}, determinations of the fine-structure constant~\cite{Bouchendira2010, Parker, Morel2020}, measurements of Newton’s gravitational constant~\cite{Rosi_2014}, and tests of the equivalence principle~\cite{PhysRevA.88.043615, PhysRevLett.113.023005}. 
See~\cite{abend2020atominterferometryapplications} for a comprehensive review.

More recently, single-photon atom interferometers based on narrow optical transitions in alkaline-earth-like atoms have been proposed for groundbreaking applications in fundamental physics~\cite{Buchmueller03042023}, such as gravitational-wave detection~\cite{PhysRevLett.110.171102, Hogan_2011, Badurina_2021} and ultra-light dark-matter searches~\cite{PhysRevLett.115.011802, PhysRevLett.117.261301, Badurina_2021}. 
These applications are examples of \emph{time-resolved measurements}, where the interferometric phase is measured repeatedly to capture dynamic signals. 
Time-resolved atom interferometry introduces unique challenges, as it demands unprecedented stability and control of systematic effects to achieve the desired sensitivity.

One of the most significant sources of systematic error in atom interferometry arises from laser wavefront aberrations, 
which cause the interferometric phase to depend on transverse atomic motion.
While the impact of wavefront aberrations has been extensively studied in the context of time-integrated atom interferometry~\cite{Wicht2005, Fils2005, Karcher2018, Louchet-Chauvet2011, Pagot2024},
where, for example, they have been identified as the main systematic in measurements of the fine-structure constant~\cite{Bouchendira2010,Parker, Morel2020}, their effects on time-resolved measurements have not been studied in detail until now. This gap in understanding is critical, as next-generation, large-scale terrestrial atom 
interferometry experiments, such as AION~\cite{Badurina_2020}, MAGIS-100~\cite{Abe_2021}, MIGA~\cite{canuel2022gravityantennabasedquantum}, ELGAR~\cite{canuel2020technologieselgarlargescale} and ZAIGA~\cite{Zhao_2022},
aim to achieve unprecedented sensitivity levels using large momentum transfer (LMT) techniques, which amplify the number of atom-light interactions, 
thereby increasing the influence of wavefront aberrations. This effect is also expected to impact future space-based atom interferometers~\cite{Hogan_2011}.

The wavefront curvature present in Gaussian beams is a simple example of 
such aberrations. In this work, we develop analytical models of the phase 
shifts induced by the coupling between the atom cloud's transverse centre-of-mass (COM) motion and wavefront 
curvature in Gaussian laser beams, and apply them to estimate 
the level of stability of the atoms’ transverse motion required to meet sensitivity goals in
100-m and 1-km atom gradiometers modelled after upcoming experiments. In doing so, we examine how the beam geometry, in particular the focus position \(f\)
and Rayleigh range \(z_R\),  influences curvature-induced noise and pulse efficiency.

We also propose a noise mitigation strategy based
on position-resolved phase-shift readout, which enables the extraction of velocity information via suitable analysis of the interferometric phase profile. This approach complements other 
mitigation strategies, such as increasing the
beam waist, by mitigating phase noise without sacrificing pulse efficiency.

The structure of this paper is as follows: In section~\ref{sec:semi-classical-modeling}, we use a semi-classical framework to derive 
expressions for the phase perturbation induced by Gaussian wavefront curvature in a symmetric LMT Mach-Zehnder atom interferometer, and validate
them using numerical simulations. In section~\ref{sec:phase-shift-measurement}, we use these results to derive the impact of 
such phase shifts on a position-averaged phase measurement protocol, validate the results using Monte Carlo simulations and apply
our findings to 100-m and 1-km atom gradiometers, highlighting the initial cloud COM position and velocity stability requirements needed to achieve phase noise levels of $10^{-5}$ rad/$\sqrt{\text{Hz}}$, the target for next-generation experiments such as AION and MAGIS-100\cite{Badurina_2020, Abe_2021}. Finally, in section~\ref{sec:mitigation}, we propose a 
mitigation strategy to reduce the noise induced by wavefront curvature in time-resolved atom interferometry experiments,
using position-resolved phase-shift readout. We then discuss the implications of our findings for the design of future atom 
interferometry experiments and highlight future work directions.

\section{Semi-Classical Modelling of the atom interferometer}\label{sec:semi-classical-modeling}
\begin{figure}[t]
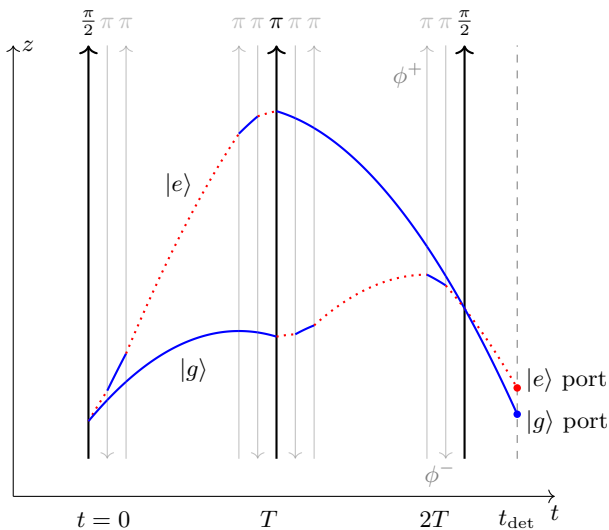

      \centering
      \include{tikz_LMT_MZ}
      \caption{Schematic of a symmetric large-momentum-transfer (LMT, $n=3$) Mach--Zehnder atom interferometer. 
Blue (solid) and red (dotted) lines denote atoms in the ground ($\ket{g}$) and excited ($\ket{e}$) states. 
The vertical axis shows position $z$ and the horizontal axis time $t$, in a uniform gravitational field. 
Vertical arrows indicate laser pulses, with duration in units of $\Omega_0^{-1}$ indicated at the top; upward and downward arrows correspond to beams propagating along $\pm z$ with spatially varying phases $\phi^{\pm}(\bm{x})$. 
Gray arrows mark the LMT pulse blocks that generate momentum separation and recombination. 
After the final beam-splitter pulse at $2T$, the two arms recombine in the output ports $\ket{g}$ and $\ket{e}$, detected at time $t_{\mathrm{det}}$.}
\label{fig:MZsequence_nlmt5}
\end{figure}

We present the framework used to model the impact of Gaussian wavefront curvature on the 
phase shift measured by an atom interferometer. Our focus is on atom interferometers based on inelastic diffraction, 
where the internal state of the atoms is modified by the atom-light interactions~\cite{Miffre2006}, and where momentum is transferred 
via single-photon transitions. For simplicity, we also restrict our attention to two-level systems.

We consider symmetric Mach-Zehnder (MZ) atom interferometers, which consist of a sequence of laser pulses that coherently split, reflect, and 
recombine atomic wave packets. The interferometer begins with an initial beam-splitter pulse, which splits the initial atom's wavefunction into a superposition of 
wave packets with different momenta. These wave packets then follow different trajectories, forming the upper and lower arms of the 
interferometer, while accumulating phase during free evolution.
After a free evolution time $T$, a mirror pulse redirects the two wave packets toward each other. Following 
another interval of free evolution of duration $T$, a final beam-splitter pulse recombines the wave packets to produce interference in the two output ports corresponding to the internal states $\ket{g}$ and $\ket{e}$. In a LMT MZ interferometer of order $n$, the two interferometer arms are separated by $n \hbar k$ and the full sequence
contains  
$4(n - 1)$ counter-propagating $\pi$-pulses in addition to the two beam-splitter pulses and one mirror pulse
in a simple MZ interferometer. Figure~\ref{fig:MZsequence_nlmt5} illustrates such a sequence for 
$n=3$. 

We use a \emph{semi-classical} (or \emph{ray atom optics}) model~\cite{MeystreTextBook, Moore1997}, in which atoms are treated as point-like particles following classical trajectories, while their internal states evolve quantum mechanically as a two-level system~\cite{Moore1997}. 
Each atom thus follows two classical trajectories (upper and lower arms) during the interferometric sequence, along which a complex phase is accumulated. We ignore the effect of parasitic paths~\cite{PhysRevLett.131.033602}.
For a given atom characterised by initial coordinates $(\bm{x}_0,\bm{v}_0)$, the total interferometric phase is 
\begin{equation}
\Delta\varphi = (\varphi_{\text{tot},u} - \varphi_{\text{tot},l}) - \frac{m(\bm{v}_u + \bm{v}_l) \cdot (\bm{x}_u - \bm{x}_l)}{2\hbar},
\label{eq:tot-phase-shift-one-wp}
\end{equation}
where the quantities on the right-hand side are evaluated along classical trajectories defined by $(\bm{x}_0,\bm{v}_0)$. Here, $\varphi_{\text{tot},u}$ and $\varphi_{\text{tot},l}$ refer to the total phase accumulated along the upper and lower arm trajectories, respectively. The second term accounts for the phase difference due 
to spatial separation between trajectories at detection time~\cite{hogan2008lightpulseatominterferometry}. 

This semi-classical approach has been successfully applied to quantify the effects of wavefront aberrations in atom 
interferometers~\cite{Pagot2024, Schkolnik_2015, Parker, Bade2018} and forms the basis of our analysis.

\subsection{Phase evolution along a single trajectory}

A light-pulse atom interferometer consists of alternating intervals of free evolution and coherent interactions 
with laser pulses. 
During free evolution, a given atomic trajectory accumulates a propagation phase given by
\begin{equation}
\varphi_{\text{prop}}(\bm{x}_{\text{init}}, \bm{x}_{\text{final}}) = -\omega_0(t_{\text{final}}-t_{\text{init}})+\frac{S_{\text{cl}}(\bm{x}_{\text{init}}, \bm{x}_{\text{final}})}{\hbar},
\label{eq:propagation-phase}
\end{equation}
where $\omega_0$ is the resonant frequency of the two-level system and $S_{\text{cl}}$ is the action evaluated along the classical trajectory 
between initial and final positions~\cite{Storey1994, PhysRevLett.67.181}.

We assume that the stimulated emission and absorption of photons are driven by a laser field with
phase
\begin{equation}
\Phi(\bm{x}, t) = \bm{k} \cdot \bm{x} - \omega t + \phi(\bm{x}),\label{eq:laser-phase}
\end{equation}
where $\bm{k}$ is the wave vector, $\omega$ is the angular frequency of the laser, and $\phi(\bm{x})$ 
captures any deviation of the laser phase from the ideal plane wave case.
Note that we assume here that this deviation is time-independent.

In the semi-classical model, we assume that the atom is sufficiently localized relative to $\phi(\bm{x})$ so that,
at the $i$-th interaction, the laser field can be approximated locally by a plane wave with phase offset $\phi(\bm{x}_i)$, where $\bm{x}_i$ is the position of the atom at that interaction. This case can be treated analytically~\cite{antoine2006, hogan2008lightpulseatominterferometry}. Under this approximation, each laser interaction at time $t_i$ contributes a phase
\begin{equation}
\varphi_{\text{laser}}(\bm{x}_i, t_i) = \pm \Phi(\bm{x}_i, t_i),
\end{equation}
and imparts a momentum of $\pm \hbar \bm{k}$ onto the atom.
The sign depends on whether the transition is from ground to excited state or vice versa.

Note that we neglect here the momentum imparted by the transverse 
phase gradient of the laser beam, which is typically much smaller than the momentum imparted by 
the longitudinal phase gradient~\cite{Pagot2024}.

The total phase acquired along a single trajectory over the course of the interferometric sequence with $N_p$ laser pulses is then given by 
the sum of the propagation and laser phases:
\begin{equation}
\varphi_{\text{tot}} = \sum_{i=0}^{N_p-1}\big[ \varphi_{\text{prop}}(\bm{x}_{i}, \bm{x}_{i+1}) + \varphi_{\text{laser}}(\bm{x}_i, t_i)\big],
\end{equation}
where $\bm{x}_{N_p}$ the position along this trajectory at detection time (after the final laser pulse).

\subsection{Phase difference in a LMT Mach-Zehnder interferometer}\label{sec:mz}
For a perfectly symmetric 
MZ sequence in a uniform gravitational field, the 
separation-induced phase vanishes and the propagation phases of the two arms cancel, so that the 
residual interferometer phase is entirely due to the laser field~\cite{PhysRevLett.67.181}. We therefore only consider the laser phase contributions in the rest of this work. 

Laser phase contributions depend on trajectory, so $\Delta\varphi$ in Eq.~\eqref{eq:tot-phase-shift-one-wp} can be expressed in terms of the initial coordinates of the atom $(\bm{x}_0, \bm{v}_0)$:
\begin{equation}
\Delta\varphi(\bm{x}_0, \bm{v}_0) = \Delta\varphi_0 + \delta\varphi(\bm{x}_0, \bm{v}_0),
\label{eq:tot-phase-decomposition}
\end{equation}
with $\Delta\varphi_0$ the phase in absence of wavefront distortions, and $\delta\varphi$ the phase perturbation from $\phi(\bm{x})$.

To obtain simple analytical models, we model the LMT pulses as instantaneous blocks at $t=0$, $T$, and $2T$, each consisting of $(n-1)$, $(2n-1)$, and $(n-1)$ $\pi$-pulses, respectively. The numerical simulations used to validate these models, however, do not make this approximation and explicitly account for the finite time separation of the pulses.
With the laser wave vector along the \( z \)-axis, we define \( \phi^{\pm}(\bm{x}) \) as the spatially-varying phase of the 
laser beam propagating in the \( \pm z \) direction.
If the transverse phase profile of the laser remains approximately constant over the vertical separation of the interferometer arms during each block, 
the total perturbation can be approximated using three effective interaction points: $\bm{x}_0$, 
and the classically propagated positions $\bm{x}_{\text{prop}}(T; \bm{x}_0, \bm{v}_0)$ and $\bm{x}_{\text{prop}}(2T; \bm{x}_0, \bm{v}_0)$, where $\bm{x}_{\text{prop}}(t;\bm{x}_0,\bm{v}_0)$ is the position of a particle with initial phase-space coordinates $(\bm{x}_0,\bm{v}_0)$ propagated classically to time $t$.
The phase perturbation for a given atom is then (for $n-1$ even): 
\begin{equation} 
    \delta\varphi(\bm{x}_0, \bm{v}_0) = \frac{n+1}{2} \delta\varphi^+(\bm{x}_0, \bm{v}_0) - \frac{n-1}{2} \delta\varphi^-(\bm{x}_0, \bm{v}_0), \label{eq:lmt-phase-shift} 
\end{equation} where 
\begin{equation} 
    \delta\varphi^{\pm}(\bm{x}_0, \bm{v}_0) = \phi^{\pm}_0 - 2\phi^{\pm}_T + \phi^{\pm}_{2T},\label{eq:deltavarphi_pm}
\end{equation}
and $\phi^{\pm}_t \equiv \phi^{\pm}(\bm{x}_{\text{prop}}(t; \bm{x}_0, \bm{v}_0))$.

In single-photon clock atom interferometry based on $^{87}$Sr, which 
will be used in future large-scale experiments like AION~\cite{Badurina_2020} and MAGIS-100~\cite{Abe_2021}, pulse durations $\tau$ are $\sim$1 ms due to the narrow linewidth of the clock transition. For large $n$, the total duration of an LMT pulse block 
becomes significant, such that the space-time area enclosed by the interferometer reduces from $n\hbar k T^2/m$ to $n\hbar k T(T-n\tau)/m$~\cite{antoine2006}. 
To account for this effect, we substitute 
\begin{equation}
    T \rightarrow T_{\text{eff}} = \sqrt{T(T-n\tau)}\label{eq:teff}
\end{equation}
in Eq.~\eqref{eq:deltavarphi_pm}. Numerical simulations
in the next section validate this approximation. 

\subsection{Phase shift due to transverse motion in a Gaussian beam}\label{subsec:gaussian-beam}
\begin{figure}
    \centering
    \usetikzlibrary{arrows.meta}

\definecolor{beamred}{RGB}{220,50,47} 
\colorlet{axisgray}{gray!60}
\colorlet{linegray}{gray!70}
\colorlet{wavegray}{gray!55}
\colorlet{mirrorgray}{gray!30}
\colorlet{cloudblue}{RoyalBlue!65}
\tikzset{
  every node/.style={font=\sffamily\scriptsize},
  >={Stealth[length=1.6mm, width=1.6mm]},
  thinline/.style={line width=0.6pt},
  midline/.style={line width=0.8pt}
}

\newcommand{\ParabolicTrajectory}[6]{%
  \pgfmathsetmacro{\xmid}{(#2+#4)/2}
  \pgfmathsetmacro{\controloffset}{abs(#4-#2)*0.25}
  \draw[->, line cap=round, #1]
    (#2,#3) .. controls (#2+\controloffset,#6) and (#4-\controloffset,#6) .. (#4,#5);
}

\newcommand{\drawWavefronts}[2]{%
  \foreach \z in {0.5,1.0,...,7} { 
    \pgfmathsetmacro{\dy}{\z-#1}
    \ifdim \dy pt=0pt\else
      \pgfmathsetmacro{\w}{\wzero*sqrt(1 + ((\dy)/\zR)^2)}
      \pgfmathsetmacro{\Rraw}{\dy*(1 + (\zR/\dy)^2)}
      \pgfmathsetmacro{\r}{abs(\Rraw)}
      \pgfmathsetmacro{\ratio}{min(\w/\r,\alphacap)}
      \pgfmathsetmacro{\alphadeg}{\ratio*57.2957795}
      \pgfmathsetmacro{\midang}{ (\Rraw>0 ? 90 : -90) }
      \pgfmathsetmacro{\astart}{\midang + \alphadeg}
      \pgfmathsetmacro{\aend}{\midang - \alphadeg}
      \pgfmathsetmacro{\yc}{\z - \Rraw}
      \pgfmathsetmacro{\opa}{0.12 + 0.55*exp(-abs(\dy)/\tau)} 
      \begin{scope}[shift={(0,\yc)}]
        \draw[#2,opacity=\opa,line width=0.45pt]
          (\astart:\r) arc[start angle=\astart, end angle=\aend, radius=\r];
      \end{scope}
    \fi
  }
}

\newcommand{\UPipeRight}[6][]{%
  \pgfmathsetmacro{\xr}{#2 + 2*(#6)}%
  \draw[-{Stealth[length=1.6mm]}, line cap=round, #1]
    (#2,#3) -- (#2,#4)
    arc[start angle=180, end angle=0, radius=#6]
    -- (\xr,#5);
}
\newcommand{\UPipeLeft}[6][]{%
  \pgfmathsetmacro{\xl}{#2 - 2*(#6)}%
  \draw[-, line cap=round, #1, thinline]
    (#2,#3) -- (#2,#4)
    arc[start angle=0, end angle=180, radius=#6]
    -- (\xl,#5);
  \draw[-{Stealth[length=1.6mm]}, thinline] (\xl,#5-0.01) -- ++(0,-0.01);
}

\begin{tikzpicture}[x=0.9cm,y=0.9cm,line cap=round,line join=round]

  \def\L{8}
  \pgfmathsetmacro{\zw}{4}
  \def\wzero{1}
  \def\zR{6.5}
  \def\zRB{2.0}
  \pgfmathsetmacro{\scale}{1/0.1}
  \def\tau{1.6}     
  \def\alphacap{0.95}
  \def\zb{0.7}
  \def\zt{0.7*\L}
  \def\cloudradius{0.1}
  \def\xshift{6}
  \def\factor{3}

  \draw[beamred, midline]
    plot[domain=0:{0.9*\L}, samples=120, variable=\y]
      ({ \wzero*sqrt(1 + ((\y-\zw)/\zR)^2) }, \y);
  \draw[beamred, midline]
    plot[domain=0:{0.9*\L}, samples=120, variable=\y]
      ({-\wzero*sqrt(1 + ((\y-\zw)/\zR)^2) }, \y);

  \draw[beamred, opacity=0.5, thinline]
    plot[domain=0:{0.9*\L}, samples=100, variable=\y]
      ({ \wzero*sqrt(1 + ((\y+\zw)/\zR)^2) }, \y);
  \draw[beamred, opacity=0.5, thinline]
    plot[domain=0:{0.9*\L}, samples=100, variable=\y]
      ({-\wzero*sqrt(1 + ((\y+\zw)/\zR)^2) }, \y);

  \drawWavefronts{\zw}{wavegray}
  \drawWavefronts{-\zw}{wavegray}

  \pgfmathsetmacro{\wmirror}{1.1*\wzero*sqrt(1 + ((0-\zw)/\zR)^2)}
  \path[pattern={Lines[angle=60,distance=3pt]},
        pattern color=black!60, draw=none, opacity=0.7]
    (-\wmirror,-0.14) rectangle (\wmirror,0);
  \draw[linegray, midline] (-\wmirror,0) -- (\wmirror,0);
  \draw[linegray, midline] (-\wmirror,-0.14) -- (\wmirror,-0.14);

  \pgfmathsetmacro{\xstart}{-2*\cloudradius}
  \pgfmathsetmacro{\xend}{\xstart + 0.25}     
  \pgfmathsetmacro{\zpeak}{\zb+2}           
  \pgfmathsetmacro{\zpeakupper}{\zt+2}      
  
  \ParabolicTrajectory{linegray}{\xstart}{\zb}{\xend}{\zb+0.3}{\zpeak}
  
  \ParabolicTrajectory{linegray}{\xstart}{\zt}{\xend}{\zt+0.3}{\zpeakupper}

  \filldraw[fill=cloudblue, draw=black!70, line width=0.4pt, fill opacity=0.85]
    (-2*\cloudradius,\zb) circle[radius=\cloudradius];
  \filldraw[fill=cloudblue, draw=black!70, line width=0.4pt, fill opacity=0.85]
    (-2*\cloudradius,\zt) circle[radius=\cloudradius];

\node[below=4pt, black!80, fill=white, inner sep=1pt] at (-1*\cloudradius,\zb) {$(\mu_{x_0},\mu_{z_0})$};

\draw[black!70,->, line width=0.7pt] (-2*\cloudradius,\zb) -- ++(0.45,0) 
  node[right,black!85,fill=white,inner sep=1pt] {$\mu_{v_{x0}}$};

\draw[black!70,->, line width=0.7pt] (-2*\cloudradius,\zb) -- ++(0,0.45) 
  node[left=2pt,black!85,fill=white,inner sep=1pt] {$\mu_{v_{z0}}$};

  \def\xaxis{1.4}
  \draw[black,<->, thinline] (\wzero+1.5,0) -- node[right=2pt,black!85] {$f$} (\wzero+1.5,\zw);
  \draw[dashed,gray,opacity=0.8] (\wzero,0)--(\wzero+1.6,0);
\draw[dashed,gray,opacity=0.8] (\wzero,\zw)--(\wzero+1.6,\zw);

  \pgfmathsetmacro{\ww}{\wzero}
  \draw[black!85,<->, thinline] (-\ww,\zw) -- (\ww,\zw)
    node[midway,above=2pt,black!85] {$2w_0$};

  \coordinate (C) at (-2*\cloudradius,\zb);
  \draw[axisgray,->, thinline] (-1.9,-0.325) -- ++(4,0) node[below] {$x$};
  \draw[axisgray,->, thinline] (-1.9,-0.325) -- ++(0,0.7) node[left] {$z$};

\draw[axisgray] (0,-0.25) -- (0,-0.4);
\node[below,black!80] at (0,-0.4) {$x=0$};
\draw[axisgray] (-1.9-0.075,0) -- (-1.9+0.075,0);
\node[left,black!80] at (-1.9,0) {$z=0$};

\draw[red,thick,->] (0,\L)--(0,\L-0.7);

\draw[thick,<->] (-1.9*\wzero,\zb)--(-1.9*\wzero,\zt);
\draw[dashed,gray,opacity=0.8] (-2.1*\wzero,\zb)--(-2*\cloudradius,\zb);
\draw[dashed,gray,opacity=0.8] (-2.1*\wzero,\zt)--(-2*\cloudradius,\zt);
\node at (-2.2*\wzero,0.5*\zt+0.5*\zb) {$L$};

\end{tikzpicture}
    \caption{
Schematic of the long-baseline atom gradiometer configuration considered in this work.
Two vertically separated Mach-Zehnder atom interferometers (top and bottom) are operated
by a common Gaussian laser beam focused at position $z = f$.
The beam originates from the top of the chamber (red arrow) and is retro-reflected by a mirror at $z = 0$,
producing counter-propagating waves with spatially varying phases $\phi^{\pm}(x)$.
The atomic ensembles (represented by blue disks), launched with initial centre-of-mass coordinates
$(\mu_{x_0}, \mu_{z_0})$ and velocities $(\mu_{v_{x0}}, \mu_{v_{z0}})$,
follow ballistic trajectories under uniform gravity.
The gradiometer baseline $L$ is defined by the initial vertical separation between the two clouds.}
    \label{fig:sketch-ai}
\end{figure}
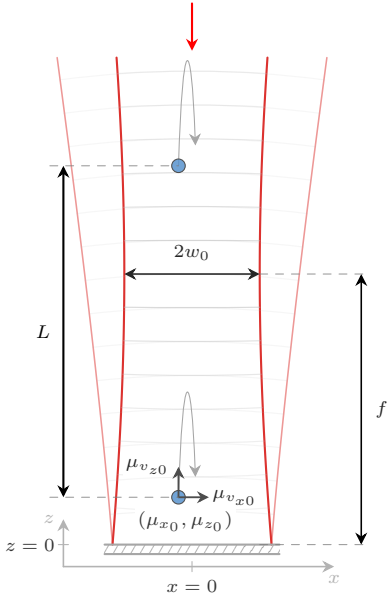

Gaussian beams provide a simple, realistic model of a perfect laser beam and represent the best possible approximation to a plane wave for a finite transverse width.  Due to the wave nature of light, they necessarily contain intrinsic wavefront curvature that persists even for perfect optical components, and can therefore introduce systematic effects under otherwise ideal conditions.

The intensity profile of a Gaussian beam is given by
\begin{equation}
    I(\bm{x}) = I_0 \left(\frac{w_0}{w(z)}\right)^2 \exp\left(-\frac{2(x^2 + y^2)}{w(z)^2}\right),\label{eq:gaussian-intensity}
\end{equation}
where \(w(z)\) is the beam radius at propagation distance \(z\), given by:
\begin{equation}
    w(z) = w_0 \sqrt{1 + \frac{z^2}{z_R^2}}.
\end{equation}
Here, \(I_0\) denotes the peak intensity at the beam focus at $z=0$ with waist $w_0$.
The position-dependent Rabi frequency \(\Omega(\bm{x})\) is given by:
\begin{equation}
    \Omega(\bm{x}) = \Omega_0 \sqrt{\frac{I(\bm{x})}{I_0}},\label{eq:rabi-freq}
\end{equation}
where \(\Omega_0 = \Omega(\bm{0})\) is the peak Rabi frequency.

Under the paraxial approximation, the phase of a Gaussian beam with wavelength $\lambda$ propagating 
along the $+z$-direction can be written as~\cite{Svelto2010}
\begin{equation}
    \Phi(\bm{x},t) = k z - \omega t + \frac{k(x^2+y^2)}{2R(z)} - \psi(z),\label{eq:gaussian-phase}
\end{equation}
where $k=2\pi/\lambda$ is the wave number, $R(z)$ is the wavefront radius of curvature,
\begin{equation}
    R(z) = z\left(1 + \frac{z_R^2}{z^2}\right),
\end{equation}
the Rayleigh range is $z_R = \pi w_0^2/\lambda$, and 
$\psi(z) = \arctan(z/z_R)$ is the Gouy phase. 

In what follows, we neglect the Gouy-phase term $\psi(z)$ in Eq.~\eqref{eq:gaussian-phase}. 
While the Gouy phase can introduce significant systematic shifts in time-integrated measurements (e.g., in determinations of the fine-structure constant~\cite{Bouchendira2010,Parker,Morel2020}) its contribution to shot-to-shot variations in the interferometric phase 
is several orders of magnitude smaller than the phase shifts induced by wavefront curvature. 
More generally, we focus on transverse effects because, whether arising through the curvature term $R(z)$ or through $\psi(z)$, longitudinal contributions remain far weaker.  
This approximation is justified in Appendix~\ref{app:long-motion}.

Under this approximation, the deviation from a plane wave (introduced in Eq.~\eqref{eq:laser-phase}) is
\begin{equation}
    \phi(\bm{x}) = \frac{k(x^2+y^2)}{2R(z)}.
\end{equation}

We consider the setup shown in Fig.~\ref{fig:sketch-ai}, where a single Gaussian beam is used to operate atom interferometers in a vertical 
vacuum chamber, with a mirror at the bottom of the chamber ($z=0$), and the laser source 
at the top. We assume that the 
beam is maximally focused at $z=f$. 
The phases of the upward and downward propagating beams are then obtained by shifting the expression for $f=0$ accordingly:
\begin{equation}
    \phi^{\pm}(\bm{x}) = \phi(x,y,f\pm z).\label{eq:phi-pm-chamber}
\end{equation}

We now focus on a single atom interferometer. Assuming atoms move over distances small compared to the Rayleigh range $z_R$, we expand $R^{-1}(f\pm z)$ around the mean initial vertical position $\mu_{z_0}$:
\begin{equation}
    R^{-1}(f\pm z) \approx R^{-1}(f\pm\mu_{z_0}) \pm (z - \mu_{z_0}) \frac{d R^{-1}}{dz}|_{z =f\pm \mu_{z_0}}.
\end{equation}
Substituting into $\phi(\bm{x})$, ignoring the Gouy term, yields:
\begin{equation}
    \phi^{\pm}(\bm{x}) \approx (c_0^{\pm} + c_1^{\pm} z)(x^2 + y^2),\label{eq:simplified-gaussian-phase}
\end{equation}
with
\begin{equation}
\begin{gathered}
    c_0^{\pm} = \frac{k\big[fz_R^2+(f\pm\mu_{z_0})^2(f\pm 2\mu_{z_0})\big]}{{2\big[z_R^2+(f\pm\mu_{z_0})^2\big]}^2},\\
    c_1^{\pm} = \pm\frac{k\big[z_R^2 - (f\pm\mu_{z_0})^2\big]}{2{\big[z_R^2+(f\pm\mu_{z_0})^2\big]}^2}.\label{eq:coefficients}
\end{gathered}
\end{equation}

\begin{figure*}
    \centering
    \includegraphics[width=\linewidth]{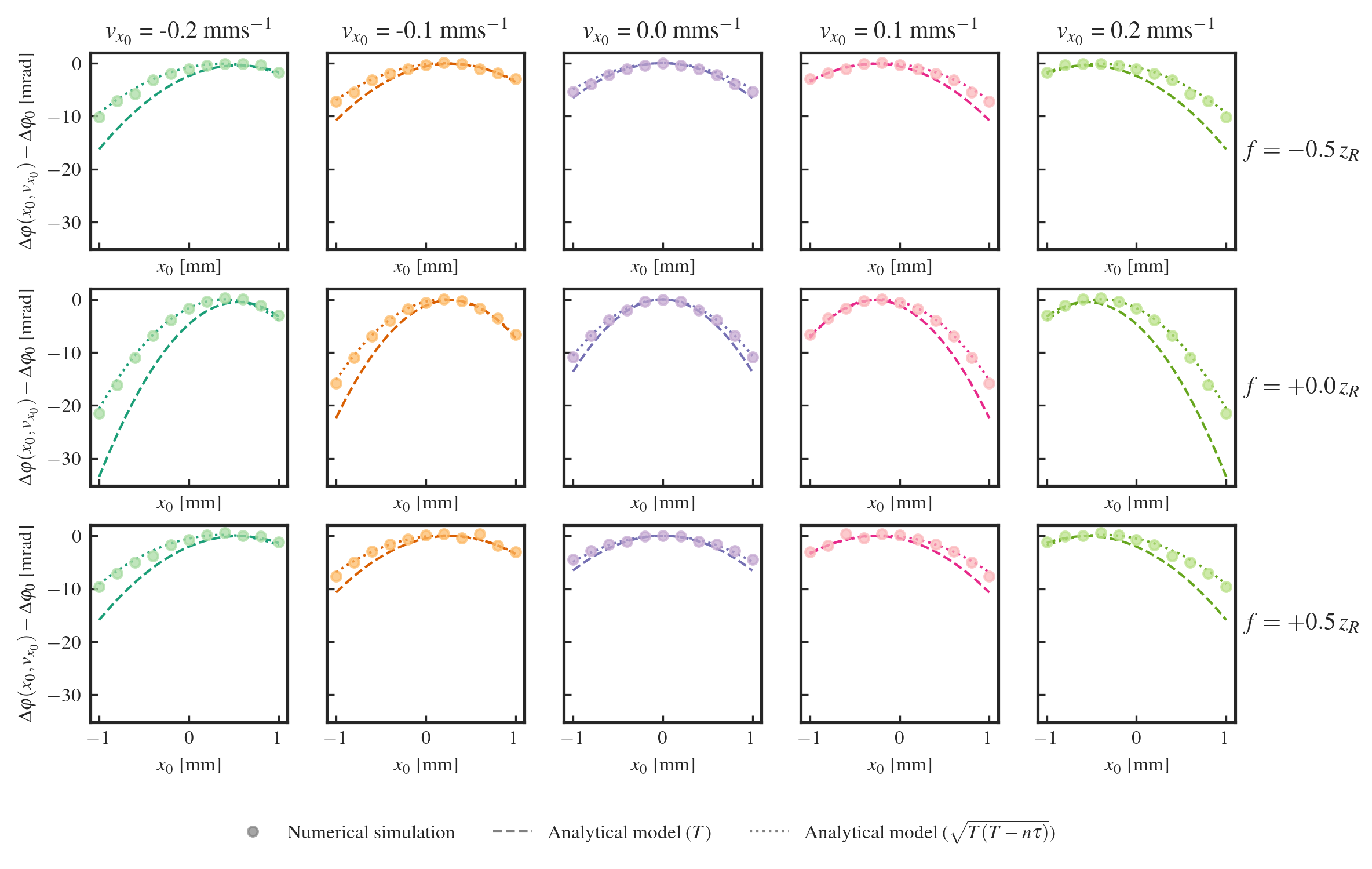}
    \caption{Phase shift for a single atom in a Gaussian beam at different focus positions $f$: comparison between numerical simulations and the analytical model of Eq.~\eqref{eq:one-atom-phase}.  
Each panel shows the phase shift $\Delta\varphi(x_0,v_{x_0})-\Delta\varphi_0$ as a function of the initial transverse position $x_0$ for several initial transverse velocities $v_{x_0}$, 
with $y_0 = v_{y_0} = 0$.  
Dashed lines correspond to the analytical model using the nominal interrogation time $T$, and dotted lines to the model with the effective interrogation time $T_{\mathrm{eff}}=\sqrt{T(T-n\tau)}$.  
Simulations (circles) were performed for a symmetric LMT Mach--Zehnder sequence on the $698$~nm clock transition of $^{87}$Sr with $n=1001$, Rabi frequency $\Omega_0=2\pi\times1$~kHz ($\tau=0.5$~ms), beam waist $w_0=3$~cm, launch velocity $v_{z_0}=19.62$~m/s, and interrogation time $T=2.225$~s ($T_{\mathrm{eff}}=1.99$~s).  
The simulations account for the finite duration of the LMT pulses, while the analytical model assumes instantaneous pulses.  
The global phase offset $\Delta\varphi_0$ is extracted from the simulation. Note that throughout the paper, the range of initial positions and velocities in simulations
            is chosen to be much larger than the typical fluctuations expected in a real experiment, to emphasize the effect of wavefront curvature.
            }\label{fig:teff-plots}
\end{figure*}

Assuming atoms evolve under uniform gravitational acceleration $g$, the position and velocity at time $t$ are given by
$\bm{x}(t)=\bm{x}_0+\bm{v}_0t+\frac{1}{2}\bm{g}t^2, \bm{v}(t)=\bm{v}_0+\bm{g}t$.
We neglect the vertical separation of the interferometer arms, as our analysis focuses on the phase response to transverse motion and shot-to-shot variations.  
For cm-scale beam waists, the Rayleigh range is on the order of hundreds of metres to a few kilometres (e.g., $z_R \approx 4$~km for $w_0 = 3$~cm in Fig.~\ref{fig:teff-plots}), 
while the arm separation ($\sim 15$~m for the parameters in Fig.~\ref{fig:teff-plots}) remains negligible in comparison, even for LMT sequences.

Using equations~\eqref{eq:phi-pm-chamber},~\eqref{eq:simplified-gaussian-phase}, and~\eqref{eq:lmt-phase-shift}, the 
wavefront-induced phase perturbation accumulated by a single atom with initial coordinates $(\bm{x}_0, \bm{v}_0)$ is:
\begin{equation}
\begin{gathered}
    \delta\varphi(\bm{x}_0,\bm{v}_0)=\frac{1}{2}T^2\bigg(c_1^+(n+1)-c_1^-(n-1)\bigg)\\\times\bigg(C^{(0)}_{xx}x_0^2+C^{(0)}_{xv}(v_{z_0})v_{x_0}x_0+C^{(0)}_{vv}(z_0,v_{z_0})v_{x_0}^2\bigg)
    \\+(\text{equivalent terms for $y$}),\label{eq:one-atom-phase}
\end{gathered}
\end{equation}
where the coefficients $C^{(0)}_{ij}$ are defined in 
Table~\ref{tab:phase-shift-coefficients}.

As further discussed in Section~\ref{sec:effect-focus-position}, the behaviour of the phase perturbation depends on the focus position $f$ and exhibits two limiting regimes of interest: $f=0$ (focus at the mirror) and $f=\pm z_R$ (focus at the Rayleigh range). When $\mu_{z_0}, L\ll z_R$, these limiting cases show distinct scalings.
At focus ($f=0$), the perturbation scales as $\delta\varphi \sim n z_R^{-2} \propto n w_0^{-4}$, whereas near the Rayleigh range ($f=\pm z_R$), it scales as $\delta\varphi \sim z_R^{-1} \propto w_0^{-2}$.
For intermediate focus positions, the $n z_R^{-2}$ dependence typically dominates in LMT sequences because of the additional scaling with $n$. 
\begin{table}[]
    \centering
    \caption{Analytical expressions for the coefficients appearing in the single-atom phase perturbation due to wavefront curvature in a Gaussian beam
    (Eq.~\ref{eq:one-atom-phase}).}\label{tab:phase-shift-coefficients}
    \begin{tabular}{c|c}
        Coefficient & Expression \\
         \hline\hline
       $C^{(0)}_{xx}$  & $-g$ \\ 
       $C^{(0)}_{xv}(v_{z_0})$ & $4v_{z0}-6gT$ \\
       $C^{(0)}_{vv}(z_0,v_{z_0})$ & $\frac{2\big(c_0^+(n+1)-c_0^-(n-1)\big)}{\big(c_1^+(n+1)-c_1^-(n-1)\big)}-(7gT^2-6Tv_{z_0}\!-\!2z_0)$
    \end{tabular}
    
\end{table}
To validate this model, we numerically simulate single-atom, $1001\hbar k$, MZ interferometers, 
driven by a Gaussian laser beam. LMT pulses are applied at regular time intervals in the simulation to account for finite duration.
The phase shift as a function of initial $x_0$ and $v_{x_0}$ is shown in Fig.\ref{fig:teff-plots}, and compared with 
the analytical prediction of Eq.~\eqref{eq:one-atom-phase}, using both the nominal interrogation time $T$ and the effective interrogation time
$T_{\text{eff}}$ defined in Eq.~\eqref{eq:teff}. The agreement between simulation and theory improves significantly when using the effective time.

\subsection{Effective phase perturbation in a position-averaged measurement}
\begin{figure*}
    \centering
    \includegraphics[width=\linewidth]{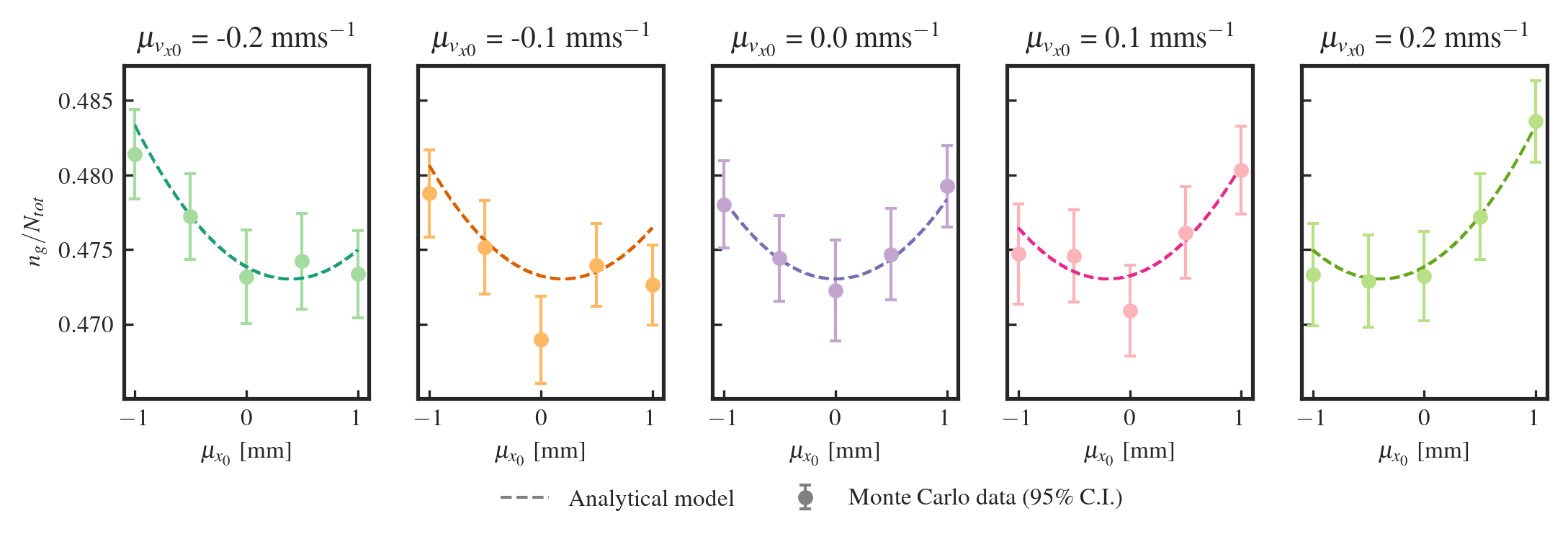}
    \caption{normalised ground-state population as a function of the initial transverse COM position and velocity in a $1001\hbar k$ MZ sequence: comparison
            between Monte Carlo simulations and our analytical model. 
            The normalised population is computed for $10^6$ initially Gaussian-distributed atoms with varying initial transverse COM position $\mu_{x_0}$ and velocity $\mu_{v_{x0}}$.
            Simulation parameters are the same as in Fig.~\ref{fig:teff-plots}.
            Dashed lines represent the analytical prediction $\frac{1}{2}(1+\cos(\Delta\varphi_0+\overline{\delta\varphi}))$, where
            $\Delta\varphi_0$ is a global phase offset obtained using the simulation results, and
            $\overline{\delta\varphi}$ is the analytical model for the
            effective phase perturbation in Eq.~\eqref{eq:pa-bias-gaussian}, using the effective interrogation time $T_{\text{eff}} = \sqrt{T(T-n\tau)}$ (Eq.~\eqref{eq:teff}), and assuming $f=0$.
            Confidence intervals are obtained 
            using bootstrapping.}\label{fig:npds}
\end{figure*}
In practice, atom interferometers operate on an ensemble of atoms characterised by 
an initial phase-space distribution $p_0(\bm{x}_0,\bm{v}_0|\bm{\theta})$, where $\bm{\theta}$ denotes experimental parameters that may vary from shot to shot due to environmental or technical noise (e.g., fluctuations in the initial atom-cloud position or velocity, or beam alignment). The interferometric phase shift modulates the probability of detecting the atom in the port corresponding to state $s$ given its initial coordinates, $p_0(s|\bm{x}_0,\bm{v}_0)$. For example, for the ground-state port $s=g$:
\begin{equation}
p_0(g|\bm{x}_0,\bm{v}_0) = \frac{1}{2}\!\left[1 + C\cos\!\big(\Delta\varphi(\bm{x}_0,\bm{v}_0)\big)\right],
\label{eq:transition-probability-pa}
\end{equation}
where $\Delta\varphi(\bm{x}_0,\bm{v}_0)$ is the trajectory-dependent phase shift derived in Sec.~\ref{sec:semi-classical-modeling}.

In position-averaged phase-shift readout, the interferometric phase shift is estimated by counting the number of atoms 
detected in each output port of the interferometer at detection time, therefore integrating over the phase-space coordinates of the detection-time joint distribution, $p(s,\bm{x},\bm{v}|\bm{\theta})$.
The expected normalised count in the ground port is given by:
\begin{equation}
\begin{aligned}
    \frac{\braket{n_g(\bm{\theta})}}{N_{\text{tot}}}\! 
    &=\! \int d\bm{x} \, d\bm{v} \, p(g, \bm{x}, \bm{v} | \bm{\theta}) 
    \!=\! \int d\bm{x}_0 \, d\bm{v}_0 \,  p_0(g, \bm{x}_0, \bm{v}_0 | \bm{\theta}) \\
    &= \int d\bm{x}_0 \, d\bm{v}_0 \, p_0(g | \bm{x}_0, \bm{v}_0) \, p_0(\bm{x}_0, \bm{v}_0 | \bm{\theta}) \\
    &=\! \frac{1}{2} \! \int \! d\bm{x}_0 d\bm{v}_0 \!\left[1 \!+\! C\! \cos\!\big(\Delta\varphi(\bm{x}_0,\! \bm{v}_0)\big)\right] \!p_0(\bm{x}_0,\! \bm{v}_0 | \bm{\theta}),
\end{aligned}
\end{equation}
where the equation in the first line uses the conservation of phase-space volume under Hamiltonian evolution~\cite{Tolman1938},
the second line uses the chain rule of probability,
and the third line substitutes Eq.~\eqref{eq:transition-probability-pa}.
We assume that all atoms are detected in either the ground 
or excited port, such that $N_{\text{tot}} = n_g + n_e$, and ignore effects due to parasitic paths or open interferometry 
ports~\cite{PhysRevLett.131.033602}.

Assuming the wavefront-induced perturbation $\delta\varphi$ is small, 
we can use the approximation
\begin{equation}
    \frac{\braket{n_g(\bm{\theta})}}{N_{\text{tot}}} \approx \frac{1}{2} \left[1 + C \cos\big(\Delta\varphi_0 + \overline{\delta\varphi}(\bm{\theta})\big) \right],
    \label{eq:exp-approx-pa}
\end{equation}
where the \emph{effective phase perturbation} in position-averaged phase-shift readout is defined as
\begin{equation}
    \overline{\delta\varphi}(\bm{\theta}) = \int d\bm{x}_0 \, d\bm{v}_0 \, \delta\varphi(\bm{x}_0, \bm{v}_0) \, p_0(\bm{x}_0, \bm{v}_0 | \bm{\theta}).
    \label{eq:pa-bias}
\end{equation}

The experimental data is then fitted to the parametric model
\begin{equation}
    \frac{n_g}{N_{\text{tot}}} = \frac{1}{2} \left(1 + C \cos(\Delta\varphi)\right),
    \label{eq:pa-fringe}
\end{equation}
to extract estimates of the phase shift $\Delta\varphi$ and contrast $C$. We emphasize that here,
$\Delta\varphi$ is a free parameter of the model (not to be confused with the phase shift
in absence of wavefront aberrations $\Delta\varphi_0$, or $\Delta\varphi(\bm{x}_0,\bm{v}_0)$).
Note that the model is not 
invertible as-is because $C$ and $\Delta\varphi$ are not linearly independent. In practice, $\Delta\varphi_0$ 
is scanned in a controlled manner to generate a set of measurement points, which are then jointly fit to 
Eq.~\eqref{eq:pa-fringe} (see example in Ref.~\cite{PhysRevLett.67.181}).

As a result, for a given set of parameters $\bm{\theta}$, the interferometer phase will be systematically biased by the effective phase perturbation:
\begin{equation}
    \Delta\varphi = \Delta\varphi_0 + \overline{\delta\varphi}(\bm{\theta}).
\end{equation}

We now compute the effective phase perturbation induced by fluctuations in the initial 
transverse COM motion of a Gaussian-distributed atom ensemble in a Gaussian beam. 
We assume the vertical COM position and velocity are fixed and exploit the cylindrical symmetry 
of the Gaussian beam to restrict our analysis to the $x$-direction.
The initial distribution is thus given by:
\begin{equation}
    p_{0}(\bm{x}_0, \bm{v}_0 | \mu_{x_0}, \mu_{v_{x0}}) = \mathcal{N}(\bm{\mu}_{x_0}, \sigma_{x_0}^2 \mathbb{I})\, \mathcal{N}(\bm{\mu}_{v_0}, \sigma_{v_0}^2 \mathbb{I}),\label{eq:gaussian-distribution}
\end{equation}
where $\bm{\mu}_{x_0} = {(\mu_{x_0}, 0, \mu_{z_0})}^T$ and $\bm{\mu}_{v_0} = {(\mu_{v_{x0}}, 0, \mu_{v_{z_0}})}^T$, and 
$\mathcal{N}(\bm{\mu}, \bm{\Sigma})$ denotes a multivariate normal distribution with mean $\bm{\mu}$ and covariance matrix $\bm{\Sigma}$~\cite{Gut1995}.
Although written in vector form for completeness, only the $x$-components are treated as varying parameters in what follows.  
Shot-to-shot variations in $\mu_{x_0}$ and $\mu_{v_{x0}}$ are thus encoded in $\bm{\theta}=(\mu_{x_0},\mu_{v_{x0}})$, 
while $\mu_{z_0}$ and $\mu_{v_{z0}}$ are held fixed.  
In practice, atom number and temperature fluctuations can also cause the spreads $\sigma_{x_0}$ and $\sigma_{v_0}$ to vary between shots.
We discuss this case in section~\ref{subsec:requirements}.

Because the perturbation in Eq.~\eqref{eq:one-atom-phase} is quadratic in $x_0$ and $v_{x_0}$,
the effective phase perturbation in position-averaged phase-shift readout can be readily computed as the average with respect to 
the Gaussian distribution in Eq.~\eqref{eq:gaussian-distribution}:
\begin{equation}
\begin{gathered}
    \overline{\delta\varphi}(\mu_{x_0}, \mu_{v_{x0}})=\frac{1}{2}T^2\bigg(c_1^+(n+1)-c^-_1(n-1)\bigg)\\
    \times\bigg(C^{(0)}_{xx}\mu_{x_0}^2+
    C^{(0)}_{xv}(\mu_{v_{z_0}})\mu_{v_{x_0}}\mu_{x_0}
    +C^{(0)}_{vv}(\mu_{z_0},\mu_{v_{z0}})\mu_{v_{x0}}^2\bigg)
    \\+(\text{constant terms.}),
\end{gathered}
\label{eq:pa-bias-gaussian}
\end{equation}
where we did not explicitly include terms that are constant in $\mu_{x_0}$ and $\mu_{v_{x0}}$.

To validate this model, we conducted Monte Carlo (MC) simulations, the 
details of which are described in Appendix~\ref{app:simulations}.
We simulated $1001\hbar k$ 
MZ sequences operated by a laser beam with Gaussian wavefront curvature (Eq.~\eqref{eq:gaussian-phase})
and Gaussian intensity profile (Eq.~\eqref{eq:gaussian-intensity}), while varying the 
initial COM position and velocity of the atom ensemble. We also set $f=0$ to increase the magnitude of the effect. Results, shown 
in Fig.~\ref{fig:npds}, are consistent with our model. 

\section{Wavefront-curvature noise in long-baseline atom gradiometers}\label{sec:phase-shift-measurement}

Atom \emph{gradiometers} operate by simultaneously measuring the phase
shift in two vertically-separated atom interferometers operated by a
common laser beam. By taking differential measurements, they suppress
common-mode noise sources such as laser phase noise~\cite{baynham2025prototypeatominterferometerdetect}, enabling
time-resolved applications that are otherwise limited.
As such, they are expected to serve as the workhorse
for next-generation precision instruments, including gravitational
wave detectors and dark matter searches~\cite{Badurina_2020, Abe_2021}.
The sensitivity of atom gradiometers to science signals typically
scales with the vertical separation, or \emph{baseline},
between the two atom interferometers ($L$ in Fig.~\ref{fig:sketch-ai}). This scaling has motivated the 
proposal of several 100-meter-scale experiments, such as MAGIS-100, 
which is already under construction, and AION-100. Larger, km-scale 
follow-up experiments have also been proposed~\cite{Abe_2021,Badurina_2020}.

Gaussian wavefront curvature coupled to transverse COM fluctuations causes phase noise in long-baseline atom gradiometers. In this section, we show that choosing a focus position $f=\pm z_R$ minimises the longitudinal variation of curvature, suppresses curvature-induced noise, and removes the $n$-scaling of LMT contributions to leading order. However, this choice also increases the difference in on-axis Rabi frequency between the two interferometer locations, degrading $\pi$-pulse efficiency. We quantify this trade-off for 100 m and 1 km baselines and derive transverse-stability requirements for two representative configurations: a low-efficiency/low-noise (LELN, $f=z_R$) and a high-efficiency/high-noise (HEHN, $f=0$) design, shown in Fig.~\ref{fig:hehn_leln_configurations}.

\subsection{Single-interferometer and gradiometer curvature noise}
\label{subsec:single-ifm}

We consider transverse COM coordinates and velocities $(\mu_{x_0},\mu_{y_0})$ and $(\mu_{v_{x0}},\mu_{v_{y0}})$ that are independent, zero-mean, Gaussian random variables with identical standard deviations $\Delta\mu_{x_0}$ and $\Delta\mu_{v_{x0}}$ in each transverse direction. (Below we also discuss the transverse spreads $\sigma_{x_0}$ and $\sigma_{v_{x0}}$ of the cloud; we reserve $\Delta\sigma$ for the shot-to-shot standard deviation of those spreads, consistent with our usage elsewhere).

Let $\hat{\Delta\varphi}$ denote the single-shot interferometric phase estimator, with variance $\sigma^2_{\hat{\Delta\varphi}}=\mathrm{Var}[\hat{\Delta\varphi}]$ over experimental realizations. Ignoring technical noise, we write $\sigma^2_{\hat{\Delta\varphi}}=\sigma^2_{\text{asn}}+\sigma^2_{\text{curv}}$, where $\sigma_{\text{asn}}$ is atom-shot noise and $\sigma_{\text{curv}}$ arises from Gaussian wavefront curvature. Shot-noise-limited operation requires $\sigma_{\text{curv}}\le\sigma_{\text{asn}}$.

The curvature-induced variance is obtained by evaluating the variance of the effective wavefront-induced phase-shift perturbation in Eq.~\eqref{eq:pa-bias-gaussian}, including a factor of two to account for both transverse directions:
\begin{equation}
    \begin{gathered}
        \sigma_{\text{curv}}^2(\mu_{z_0})=\frac{1}{2}T^4\bigg(c_1^+(n+1)-c^-_1(n-1)\bigg)^2
    \bigg(2C^{(0)2}_{xx}\Delta\mu_{x_0}^4\\+
    C^{(0)2}_{xv}(\mu_{v_{z0}})\Delta\mu_{v_{x0}}^2\Delta\mu_{x_0}^2
    +2C^{(0)2}_{vv}(\mu_{z_0},\mu_{v_{z0}})\Delta\mu_{v_{x0}}^4\bigg).\label{eq:pa-var}
    \end{gathered}
\end{equation}
where $c_1^\pm$ and $C^{(0)}_{ij}$ are defined in Table~\ref{tab:phase-shift-coefficients}.

A gradiometer consists of two identical interferometers separated vertically by a baseline $L$. Assuming uncorrelated transverse COM fluctuations, the total curvature-induced variance is simply the sum of the single-interferometer variances:
\begin{equation}
\sigma^2_{\text{grad}} = \sigma^2_{\text{curv}}(0) + \sigma^2_{\text{curv}}(L).
\label{eq:grad-uncorr}
\end{equation}
We refer to $\sigma_{\text{grad}}$ as the wavefront-curvature-induced gradiometer noise.

\subsection{Effect of the focus position $f$}\label{sec:effect-focus-position}
\subsubsection{Curvature variation along $z$ and noise suppression}
\begin{figure}
    \centering
    \includegraphics[width=0.9\linewidth]{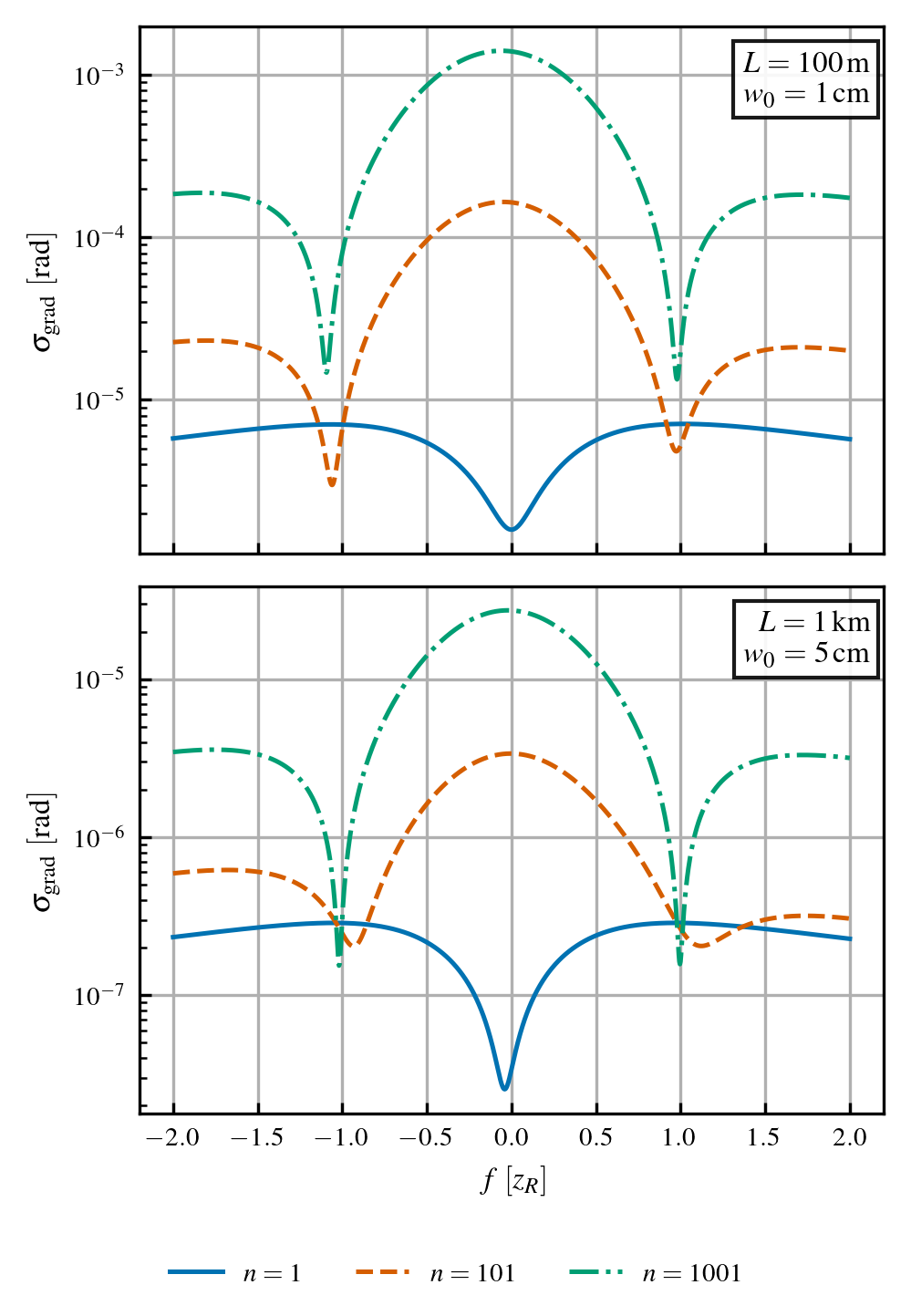}
    \caption{Curvature-induced differential phase noise in position-averaged readout versus focus position $f$, and assuming fixed COM shot-to-shot fluctuations: $\Delta\mu_{x_0}=10$ $\mu$m, $\Delta\mu_{v_{x0}}=10$ $\mu$ms$^{-1}$. Interrogation time and initial launch velocity as in Fig.~\ref{fig:teff-plots}.}
    \label{fig:noise_vs_f}
\end{figure}

For a Gaussian beam, the wavefront curvature is given by  $1/R(z)=z/(z^2+z_R^2)$. Its longitudinal derivative
\begin{equation}
\frac{\mathrm{d}}{\mathrm{d}z}\left(\frac{1}{R(z)}\right)=\frac{z_R^2 - z^2}{\bigl(z^2+z_R^2\bigr)^2}
\label{eq:d1overR}
\end{equation}
vanishes at $|z|=z_R$ and is maximal at $z=0$. Thus, choosing the focus position $f$, in Fig.~\ref{fig:sketch-ai}, so that the atom trajectories lie near the minima of Eq.~\eqref{eq:d1overR} makes the curvature locally constant, which causes the LMT wavefront contributions to the phase shift to cancel to leading order (formally, $\delta\varphi^+\approx\delta\varphi^-$ in Eq.~\eqref{eq:lmt-phase-shift}). In this limit, curvature-induced noise loses its $n$-scaling and, to leading order, only fluctuations in COM velocity contribute. 
Figure \ref{fig:noise_vs_f} shows the dependence of the gradiometer noise (Eq.~\eqref{eq:grad-uncorr}) on $f$, clearly exhibiting the minima around $|f|=z_R$, and the maximum near $f=0$. Plots suggest that, for example, for $n=1001$, a noise suppression of order 100 is possible. 

\subsubsection{Impact on LMT $\pi$-pulse efficiencies}
\begin{figure*}
\centering
\includegraphics[width=0.98\linewidth]{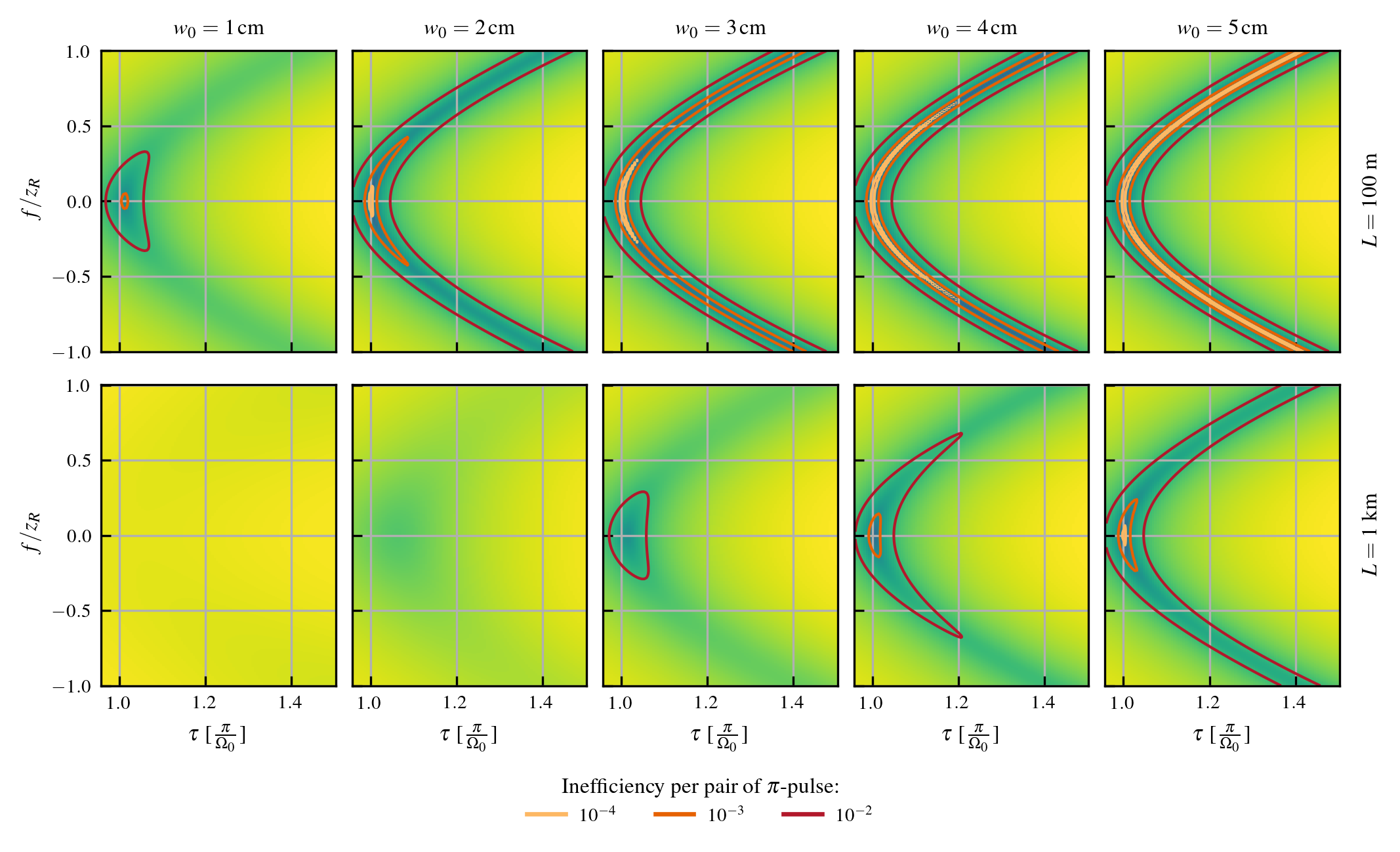}
\caption{Gradiometer inefficiency $1-\varepsilon_{\text{grad}}$ for the configuration illustrated in Fig.~\ref{fig:sketch-ai}, shown as a function of the focus position $f$, beam waist $w_0$, and pulse duration $\tau$ for 100-m (top) and 1-km (bottom) setups.
The gradiometer efficiency $\varepsilon_{\text{grad}}$ is defined in Eq.~\eqref{eq:grad-efficiency}.
Minima occur near $f=0$, coinciding with the curvature-noise maxima shown in Fig.~\ref{fig:noise_vs_f}.
Contours at $10^{-2}$, $10^{-3}$, and $10^{-4}$ mark regions where efficient common-mode operation of gradiometers with $n=10^2$, $10^3$, and $10^4$ is achievable, respectively.}
\label{fig:grad-inefficiencies}
\end{figure*}
We evaluate the average $\pi$-pulse efficiency across both interferometer locations as a function of the beam waist $w_0$, the focus position $f$, and the pulse duration $\tau$. The on-axis Rabi frequency is $\Omega_0(z)=\Omega(x{=}0,y{=}0,z)$, where $\Omega(\bm{x})$ is given in Eq.~\eqref{eq:rabi-freq}. In our setup, the efficiency of a resonant square pulse propagating along the $\pm z$ direction at position $z$ is
\begin{equation}
\varepsilon^{\pm}(z,f,\tau^{\pm}) = \sin^2\left(\frac{\Omega_0(f\pm z)\tau^{\pm}}{2}\right),
\label{eq:pulse-efficiency}
\end{equation}
and the average efficiency for the two interferometers separated by a baseline $L$ per LMT pulse pair is defined as
\begin{equation}
\begin{gathered}
\varepsilon_{\text{grad}}(f,\tau^+,\tau^-,L) = \frac{1}{2}\Big[\varepsilon^+(0,f,\tau^+)\varepsilon^-(0,f,\tau^-) \\
+ \varepsilon^+(L,f,\tau^+)\varepsilon^-(L,f,\tau^-)\Big].\label{eq:grad-efficiency}
\end{gathered}
\end{equation}
For an LMT sequence of order $n$ with per-pair-of-pulse efficiency $\varepsilon_{\text{grad}}$, the total survival fraction is $(1-\varepsilon_{\text{grad}})^{2n}\approx e^{-2n\varepsilon_{\text{grad}}}$, so maintaining $\gtrsim e^{-1}$ atoms requires $\varepsilon_{\text{grad}}\lesssim1/(2n)$, e.g. $\mathcal{O}(10^{-3})$ for $n\sim10^3$.

Figure~\ref{fig:grad-inefficiencies} shows the gradiometer inefficiency $1-\varepsilon_{\text{grad}}$ versus $f$, $w_0$, and $\tau$ for 100-m and 1-km baselines. The calculations assume $\tau^+=\tau^-=\tau$ for simplicity, though relaxing this assumption leads to the same qualitative conclusions.

The most power-efficient configuration is $f=0$, which minimises the required waist size for a given target gradiometer inefficiency. The figure also identifies combinations of $f$, $w_0$, and $\tau$ that enable efficient LMT operation. For instance, for $w_0=1$ cm and $L=100$ m, there is no pulse duration $\tau$ that yields an inefficiency as low as $10^{-3}$ unless $f\approx0$, regardless of available laser power.

This is because both interferometers are driven by the same beam, so a single pulse duration must provide efficient population transfers at both locations. This constrains the allowable variation of the Rabi frequency along the baseline. Since the rate of change of $w(z)$ is zero at the focus and becomes constant at $z\to\pm\infty$, placing the focus at $z=0$, or equivalently ensuring that both interferometers lie within the slowly varying region of the beam waist (small $L/z_R$), satisfies the efficiency condition. Alternatively, pulses that are robust to pulse-area errors, such as composite pulses~\cite{Genov2014}, could be used to mitigate these limitations.

For a fixed baseline $L$, maintaining common-mode LMT operation therefore requires larger $w_0$ when using $f=z_R$ compared to $f=0$, and thus greater laser power to preserve the same Rabi frequency. For example, from Fig.~\ref{fig:grad-inefficiencies}, for $L=100$ m, reaching $1-\varepsilon_{\text{grad}}=10^{-3}$ requires $w_0=1$ cm for $f=0$ but $w_0=3$ cm for $|f|=z_R$. Because of the defocusing, the Rabi frequency at $z=0$ furthermore decreases by a factor of $\sqrt{2}$ compared to the focus at $|f|=z_R$, which, together with the $3^2$ scaling from the waist increase, results in an overall $18$-fold increase in required laser power.

Thus, although configurations with $|f|=z_R$ exhibit lower curvature-induced phase noise, operating near $f=0$ offers clear practical advantages by minimizing power demands while maintaining efficient LMT sequences.

\subsection{Transverse motion requirements for next-generation long-baseline gradiometers}
\label{subsec:requirements}

\begin{figure*}
    \centering
    \include{tikz_gaussian_beam_HEHN_LELN}
    \caption{Schematics of the (a) high-efficiency/high-noise (HEHN, $f=0$) and (b) low-efficiency/low-noise (LELN, $f=z_R$) gradiometer configurations. Within each panel, left: Gaussian beam geometry showing two vertically separated atom clouds (blue circles, separated by baseline $L$), with the beam originating from above and retro-reflected by a mirror at $z=0$. Dashed horizontal lines indicate the focus position $f$. Right: Inverse wavefront curvature radius $1/R(z)$ for the incident ($1/R^-(z)$, solid black) and reflected ($1/R^+(z)$, dotted black) beams. White circles mark the focus positions ($z=f$); filled black circles mark curvature extrema at $z=f\pm z_R$. The HEHN configuration (a) places the focus at the mirror ($f=0$), minimizing beam waist and Rabi frequency variation across the baseline, ensuring high pulse efficiency at both interferometer locations. However, the curvature mismatch between incident and reflected beams maximises sensitivity to transverse motion, resulting in high phase noise. The LELN configuration (b) positions the focus at $f=z_R$, placing both atom clouds near the curvature extrema, where $1/R^-(z)$ is approximately constant and matches $1/R^+(z)$. This suppresses phase noise from transverse motion, but the increased beam waist variation across the baseline reduces pulse efficiency.}
\label{fig:hehn_leln_configurations}
\end{figure*}

\begin{figure}
\centering
\includegraphics[width=0.92\linewidth]{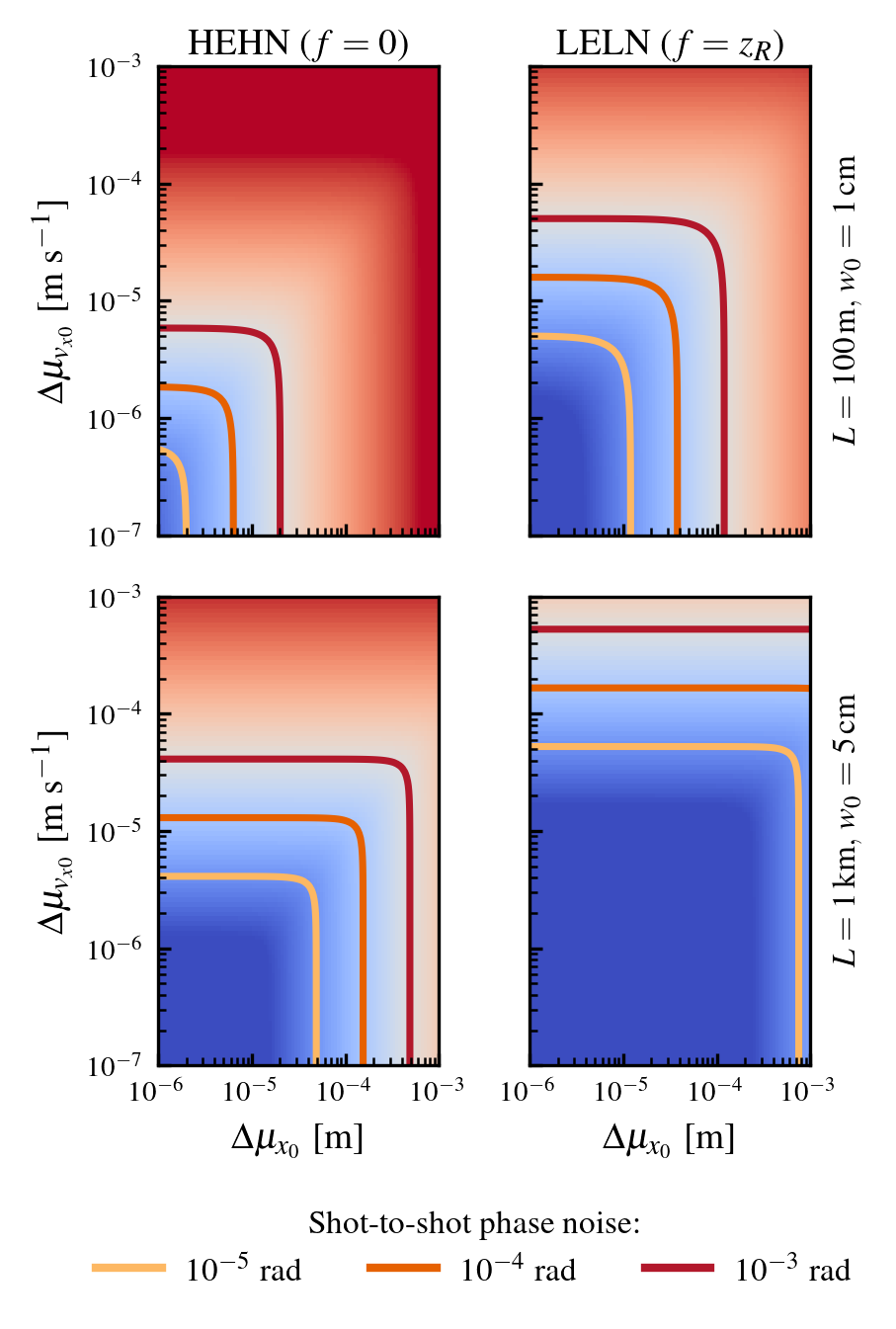}
\caption{Curvature-induced differential phase noise $\sigma_{\text{grad}}$ in position-averaged readout versus shot-to-shot COM fluctuations $\Delta\mu_{x_0}$ and $\Delta\mu_{v_{x_0}}$ for 100-m (top) and 1-km (bottom) gradiometers. Left and right columns show HEHN ($f=0$) and LELN ($f=z_R$) configurations, respectively (see Fig.~\ref{fig:hehn_leln_configurations}). Solid contours mark $\sigma_{\text{grad}}=10^{-3},10^{-4},10^{-5}$ rad. Parameters: $^{87}$Sr ($\lambda=698$ nm), LMT order $n=1001$, launch velocity and interrogation time as in Sec.~\ref{subsec:gaussian-beam}.}
\label{fig:contour-pa-grad}
\end{figure}

We have identified two limiting focus configurations: the low-efficiency/low-noise (LELN, $f=z_R$) configuration, where curvature-induced noise is minimized at the cost of Rabi-frequency uniformity, and the high-efficiency/high-noise (HEHN, $f=0$) configuration, which maximises pulse efficiency but increases sensitivity to transverse motion. 
These are represented in Fig.~\ref{fig:hehn_leln_configurations}.
In practice, next-generation detectors will likely operate in an intermediate regime, balancing both effects. 

Figure~\ref{fig:contour-pa-grad}  shows the curvature-induced differential phase noise for LELN and HEHN configurations as a function of the shot-to-shot transverse COM fluctuations $(\Delta\mu_{x_0},\Delta\mu_{v_{x0}})$ for 100-m and 1-km baselines. The contours directly map allowable transverse fluctuation levels for specified phase-noise targets, assuming $n=1001$. Note that the analysis assumes uncorrelated (shot-to-shot) Gaussian fluctuations of the initial parameters, which leads to white phase noise across all frequencies.

Achieving phase noise levels of $10^{-5}$ rad, the target for future detectors such as AION~\cite{Badurina_2020} and MAGIS-100~\cite{Abe_2021}, imposes stringent requirements on transverse motion stability. 
For a 100-m instrument with $w_0=1$ cm and $n=1001$, operation in the HEHN configuration requires transverse COM fluctuations to be stabilized at the few-micron level in position and below the micron-per-second level in velocity. In the LELN configuration, the corresponding stability requirements relax to the tens-of-microns scale in position and to just under ten microns per second in velocity. These values reflect approximately an order-of-magnitude difference between the two limits, consistent with the expected trade-off between efficiency and curvature noise suppression observed in Fig.~\ref{fig:noise_vs_f}.

By comparison, state-of-the-art atom interferometers have achieved COM position and velocities at the tens-of-microns and tens-of-microns-per-second levels, respectively~\cite{PhysRevA.101.033606,Asenbaum_2020}. Thus, meeting the HEHN requirements will require roughly a tenfold improvement over current experimental performance, whereas the LELN stability targets are already within reach, provided composite or robust pulse techniques~\cite{Genov2014} are employed to mitigate the reduced efficiency.

Note that, however, reaching these targets could be challenging for fermionic isotopes such as $^{87}$Sr, proposed 
for use in both AION and MAGIS-100, which will necessitate extremely weak trapping potentials. 
For instance, obtaining a trapped Fermi gas with $N=10^6$ atoms, while maintaining a characteristic position radius $x_F\sim\sqrt{\hbar N^{1/3}/m\omega}\sim$1 mm
and velocity radius $v_F\sim\sqrt{\hbar N^{1/3}\omega/m}\lesssim$1 mms$^{-1}$, requires significantly weaker trap frequencies on the order of \(\omega \sim 2\pi \times (0.1\,\text{Hz} - 1\,\text{Hz})\)~\cite{Dean2018}.

While the discussion above focuses on the centre-of-mass motion, the same analysis applies to the internal cloud spreads $(\sigma_{x_0},\sigma_{v_{x0}})$, which must also be stabilized at comparable levels (micron-scale in position and sub-micron-per-second in velocity) to maintain phase noise below $10^{-5}$~rad for 100-m baselines in the HEHN regime. Achieving this level of control 
over transverse spreads is expected to be similarly challenging, due to the same constraints that 
limit COM stability, i.e., the need for large atom numbers and weak trapping potentials.

Finally, we note that alternative interferometer geometries, such as multi-loop Mach-Zehnder sequences, might offer a promising avenue for suppressing curvature-induced phase shifts. However, preliminary simulations suggests that finite pulse durations limit the practical benefits of such approaches while introducing trade-offs in effective LMT order. The degree of suppression appears highly sensitive to sequence parameters, and a comprehensive optimization of such approaches is left for future work.

\subsection{Comparison to other noise sources.}
While the focus of this paper is on the effect of Gaussian wavefront curvature, 
we note that higher-order aberrations, such as those arising from imperfections 
in optical components, clipping at mirror edges, or reflections from the walls 
of the interferometry chamber, can introduce additional phase noise at higher 
spatial frequencies. These aberrations could potentially impose even more 
stringent requirements on transverse motion stability. However, their detailed 
characterisation depends on the specific optical setup used in a given experiment 
and is therefore left for future work.

In addition, terrestrial atom interferometers are subject to phase shifts 
induced by the Coriolis force due to Earth's rotation. This effect is linearly 
sensitive to initial transverse velocity 
fluctuations~\cite{PhysRevLett.111.083001}, 
and typically dominates over the curvature-induced phase noise 
discussed here, which depends quadratically on the initial position 
and velocity. Consequently, Coriolis-induced noise would generally impose even 
stricter requirements on transverse velocity stability than those derived 
above. However, future long-baseline atom interferometry 
applications, whether time-resolved or time-integrated, are expected to 
implement established Coriolis compensation 
techniques~\cite{Lan2012,glick2023coriolisforcecompensationlaser}. 
The influence of residual couplings due to imperfect compensation could 
be the subject of future investigation.

\section{Mitigation strategy: position-resolved phase-shift readout}\label{sec:mitigation}
\begin{figure*}[]
    \centering
    \includegraphics[width=\linewidth]{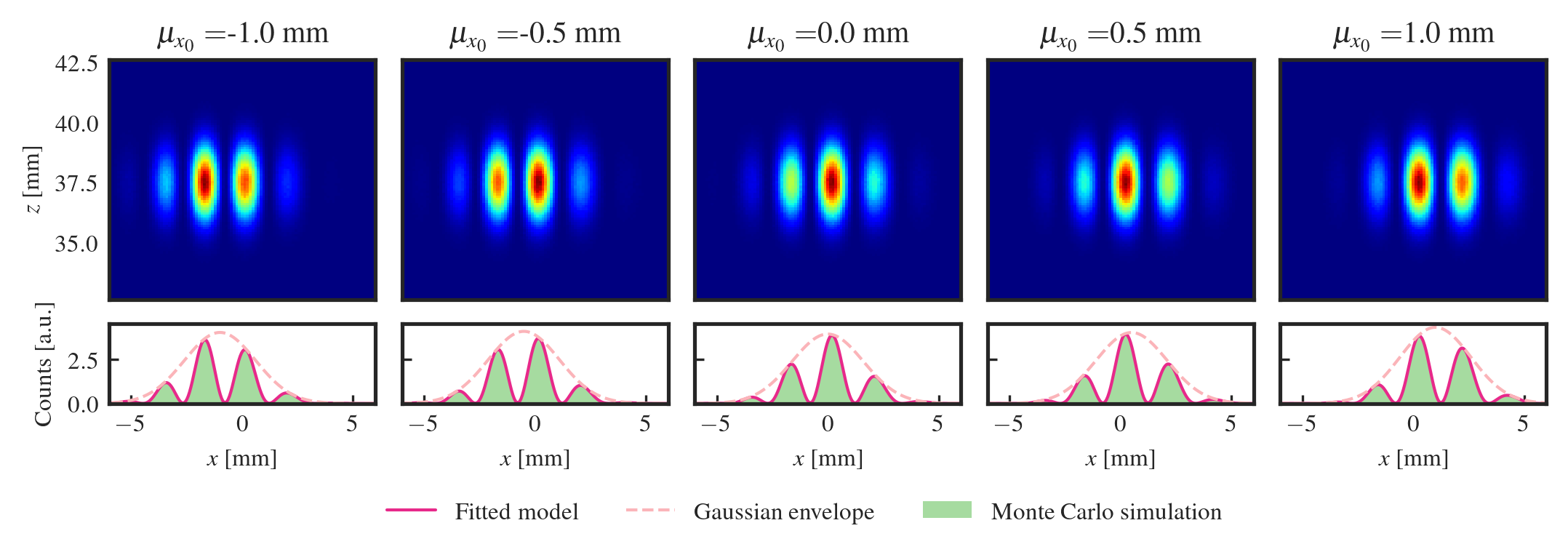}
    \caption{Top: Ground-state port atom density for simulated $101\hbar k$ LMT sequences ($N=10^7$, $\mu_{v_x}=0$) in a phase-shear readout measurement. 
    Each panel corresponds to a different initial
 COM position $\mu_{x_0}$, while the initial position and velocity spreads are kept fixed at $\sigma_{x_0}=100$ $\mu$m and $\sigma_{v_{x0}}=0.31$ mm$s^{-1}$
 (corresponding to a temperature of approximately $1$ nK).
 All the simulation parameters, apart from the LMT order $n$ and the beam 
 waist radius ($w_0=$1 cm here, equivalent to $z_R\approx450$ m) are the same as in the simulations
 of the position-averaged measurements in Fig.~\ref{fig:npds}. The simulations here also use a uniform (instead of Gaussian) intensity profile. Bottom: 1D atom densities obtained by integration along the $y$ and $z$ direction. The
 solid lines show the fit while the dashed lines show the Gaussian envelope. The extracted interferometric phase shifts 
 and phase-shift gradients are compared to our analytical model in Fig.~\ref{fig:phase-shear-phi-kappa}.}\label{fig:phase-shear-fits}
\end{figure*}

\begin{figure*}
    \centering
    \includegraphics[width=\linewidth]{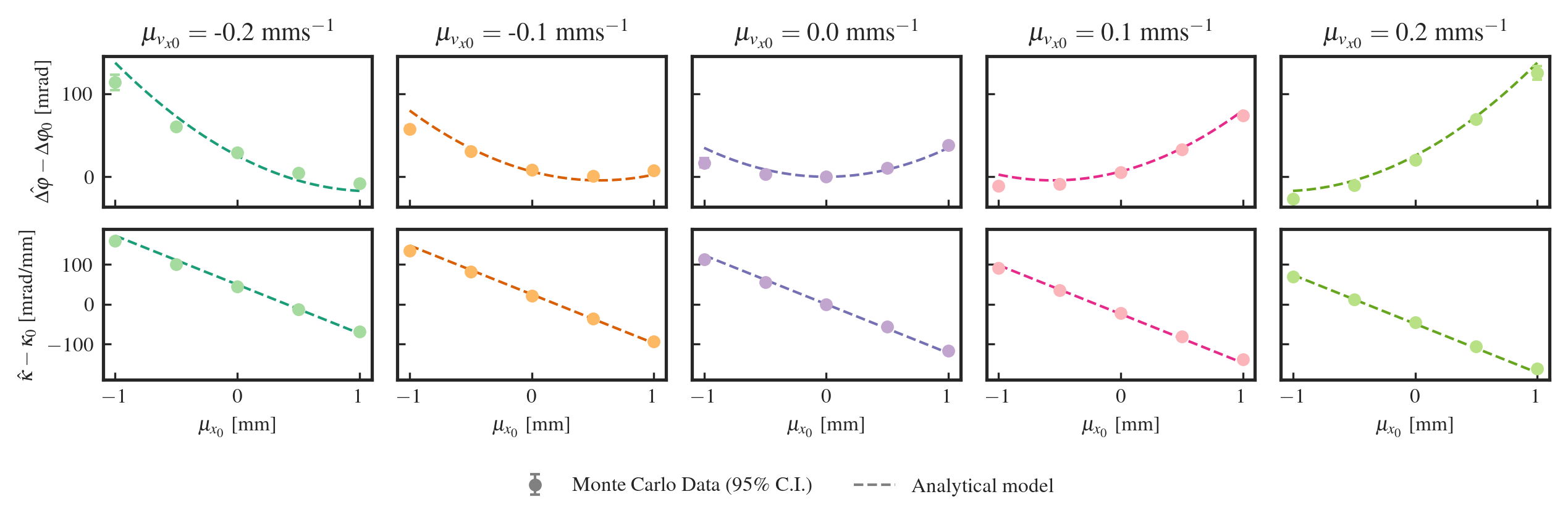}
    \caption{Comparison between Monte Carlo simulations and the analytical model in Eq.~\eqref{eq:linearized-params} for phase-shift $\Delta\varphi$ and phase-shift gradient $\kappa$ variations
    induced by transverse motion in a curved wavefront in a $101\hbar k$ LMT MZ sequence. Parameter estimates $\hat{\Delta\varphi}$ and $\hat{\kappa}$ are obtained by fitting the model in Eq.~\eqref{eq:pr-model}
    to 1D densities like the ones shown in Fig.~\ref{fig:phase-shear-fits}. Values for the unperturbed parameters, $\Delta\varphi_0$ and $\kappa_0$ are the estimates for $\mu_{x_0}=\mu_{v_{x0}}=0$.
    Confidence intervals are computed using bootstrapping.
    Note that here the parameter values are plotted as a function of the initial coordinates $(\mu_{x_0},\mu_{v_{x0}})$.}\label{fig:phase-shear-phi-kappa}
\end{figure*}

In the previous section, we assumed that all spatial information about the atom cloud was integrated over. 
Position-resolved phase-shift readout, however, has been identified as a 
promising mitigation strategy for systematic effects caused by 
wavefront aberrations in atom 
interferometry~\cite{Abe_2021,seckmeyer2024principalcomponentanalysisspatial}.
The non-uniform wavefront makes the interferometric phase sensitive to the atoms’ transverse position and velocity, resulting in phase noise when these quantities are not measured. 
While position-resolved measurements directly provide position information, velocity can also be inferred indirectly 
through suitable analysis of the position-resolved phase.
This principle is, for example, well established in point-source atom interferometry, 
where ballistic expansion creates a deterministic correlation between 
an atom’s final position and its initial velocity, allowing 
velocity-dependent phase shifts to be mapped onto the spatial 
distribution of the 
cloud at detection time~\cite{PhysRevLett.111.083001,Chen2020,PhysRevA.102.013326}.

Despite this, a detailed analysis of the benefits 
of position-resolved readout for mitigating wavefront-related systematic effects 
remained to be done. Here, we show how both position- and velocity-dependent phase shifts induced 
by curved wavefronts can be recovered from position-resolved data alone. 

\subsection{Effective phase perturbation in a position-resolved measurement}
We consider a setting in which position is resolved along the \( x \)-axis, while all other coordinates are integrated out. 
For ground-state detection, the expected (normalised) measurement outcome is given by
\begin{equation}
\begin{aligned}
    \frac{\braket{n_g(x, \bm{\theta})}}{N_{\text{tot}}} &= \int p(\bm{x}, \bm{v} | \bm{\theta}) \, p(g | \bm{x}, \bm{v}) \, d\bm{v} \, dy \, dz \\
    &= p(x | \bm{\theta}) \int p(y, z, \bm{v} | x, \bm{\theta}) \, p(g | \bm{x}, \bm{v}) \, d\bm{v} \, dy \, dz,
    \label{eq:o-psr}
\end{aligned}
\end{equation}
where we factor the distribution of classical trajectories as 
\[
p(\bm{x}, \bm{v} \mid \bm{\theta}) = p(x \mid \bm{\theta}) \, p(y, z, \bm{v} \mid x, \bm{\theta}).
\]
Here, \( p(x \mid \bm{\theta}) \) denotes the marginal spatial density 
along the resolved axis, and \( p(y, z, \bm{v} \mid x, \bm{\theta}) \) 
encodes the conditional distribution of velocities and unresolved coordinates.

Using Eq.~\eqref{eq:tot-phase-decomposition} and assuming that the perturbation $\delta\varphi$ is small, we approximate:
\begin{equation}
    \frac{\braket{n_g(x, \bm{\theta})}}{N_{\text{tot}}} \approx \frac{p(x | \bm{\theta})}{2} \left[1 + C \cos\big(\Delta\varphi_0 + \overline{\delta\varphi}(x, \bm{\theta})\big)\right],
    \label{eq:pr-data}
\end{equation}
where we defined the spatially-dependent average phase perturbation:
\begin{equation}
\begin{aligned}
    \overline{\delta\varphi}(x, \bm{\theta})\! =\! \int & d\bm{v} \, dy \, dz \, \delta\varphi\big(x^{-1}_{\text{prop}}(t_{\text{det}};\bm{x}, \bm{v}), v^{-1}_{\text{prop}}(t_{\text{det}};\bm{x}, \bm{v})\big) \\
    & \times p(y, z, \bm{v} | \bm{x}, \bm{\theta}),
    \label{eq:pr-bias}
\end{aligned}
\end{equation}
with $(x_{\mathrm{prop}}^{-1},v_{\mathrm{prop}}^{-1})$ denoting the initial position and velocity that evolve to $(\bm{x},\bm{v})$ at the detection time $t_{\mathrm{det}}$ under classical propagation.

\subsection{Learning the phase perturbation from position-resolved data}
\begin{figure}
    \centering
    \includegraphics[width=\linewidth]{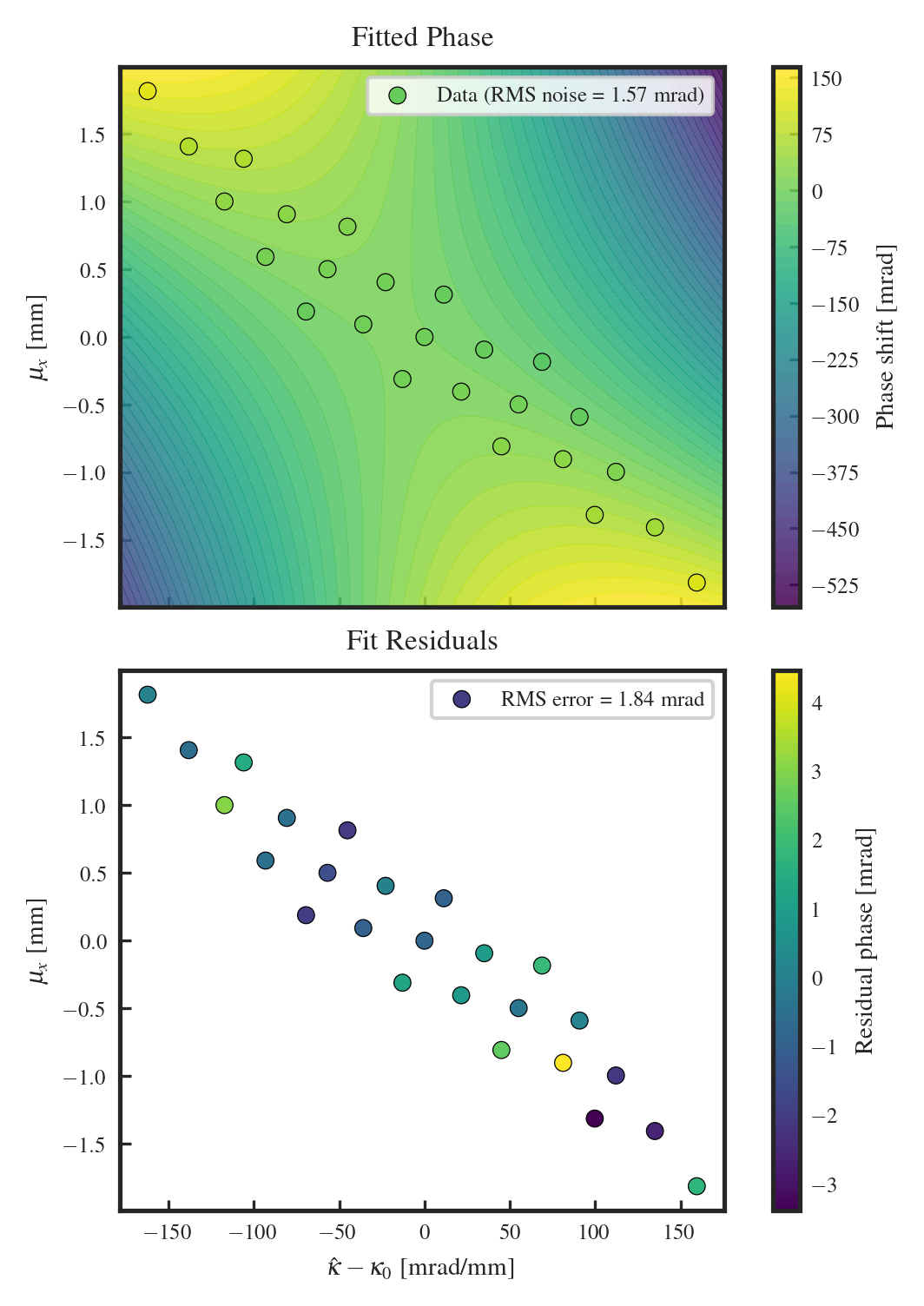}
    \caption{
    Dependence of the wavefront-curvature-induced phase-shift perturbation
    on the final COM position $\mu_x$ and phase-shift gradient $\kappa$, 
    as learned from simulated data. 
    The fitted surface has a saddle-shaped dependence on $\kappa,\mu_x$
    and is obtained from 25 MC simulations, where $\mu_x$ and $\kappa$
    extracted from histograms like those in Fig.~\ref{fig:phase-shear-fits}.
    The root-mean-square (RMS) phase noise (due to atom shot noise) and 
    the RMS fit error are indicated. Top: fitted perturbation surface 
    with data points overlaid. Bottom: fit residuals. The baseline phase shift gradient is $\kappa_0=3133$ mrad/mm.}\label{fig:fitted-phase-perturbation}
\end{figure}
 \begin{table}
    \caption{Coefficients for the position-dependent phase-shift bias in position-resolved phase-shift readout.
 We defined $C_0=\frac{1}{2}T^2[(c_1^+(n+1)-c_1^-(n-1)]$, 
 $C_{xx}=C_{xx}^{(0)}$, $C_{xv}=C_{xv}^{(0)}-2C_{xx}^{(0)}t_{\text{det}}$,  
 $C_{vv}=C_{vv}^{(0)}-C_{xv}^{(0)}t_{\text{det}}+C_{xx}^{(0)}t_{\text{det}}^2$,
and the ballistic expansion coefficient $c_{\text{exp}}=\sigma_v^2t_{\text{det}}/\sigma_x^2$,
 where $t_{\text{det}}\geq2T$ is the detection time (defined in Fig.~\ref{fig:MZsequence_nlmt5}).}
 \begin{tabular}{c|c}
    Coefficient & Expression \\
    \hline
    \hline
    $\beta_0(\mu_x,\mu_{v_x})$ & $C_0C_{vv}(\mu_{v_x}-\mu_xc_{\text{exp}})^2$\\
    $\beta_1(\mu_x,\mu_{v_x})$ & $C_0(2C_{vv}c_{\text{exp}}+C_{xv})(\mu_{v_x}-\mu_xc_{\text{exp}})$\\
    $\beta_2(\mu_x,\mu_{v_x})$ & $C_0(C_{xx}+c_{\text{exp}}(C_{vv}c_{\text{exp}}+C_{xv}))$   
 \end{tabular}\label{tab:phase-shift-coefficients-pr}

 \end{table}
An example of a position-resolved measurement protocol is phase-shear readout~\cite{Sugarbaker_2013}, in which a linear phase gradient is imprinted 
onto the atom cloud using a tilted mirror, creating spatial interference fringes whose offset encodes the interferometric phase. 
This allows for single-shot phase estimation and makes phase-shear readout
especially well suited for time-resolved atom interferometry~\cite{Abe_2021}.
In what follows, we focus on this scheme because it requires only one simulation per phase-shift estimate, 
but our conclusions could in principle apply to other measurement protocols that resolve final atomic positions.

Because the atoms experience phase curvature from the Gaussian beam, the spatially-dependent average phase 
perturbation is quadratic in $x$:
\begin{equation}
\overline{\delta\varphi}(x,\bm{\theta})=\beta_{0}(\bm{\theta})+\beta_{1}(\bm{\theta})x+\beta_{2}(\bm{\theta})x^2,\label{eq:perturbation-betas}
\end{equation}
where expressions for the coefficients $\beta_i$ are given in Table~\ref{tab:phase-shift-coefficients-pr}
and derived in Appendix~\ref{app:bias-gaussian-beam}, for the case $\bm{\theta}=(\mu_x,\mu_{v_x})^T$ (expressions in terms of initial
COM coordinates are readily obtained by substituting $\mu_x=\mu_{x_0}+\mu_{v_{x0}}t_{\text{det}}$ and $\mu_{v_x}=\mu_{v_{x0}}$).
However, if the cloud is localized around $\mu_x$, the position-resolved phase will appear approximately linear
in $x$, such that, assuming that the cloud is Gaussian-distributed in the transverse directions at detection time,
the distribution can be accurately fit to the model~\cite{Sugarbaker_2013}
\begin{equation}
    \frac{n_g(x)}{N_{\text{tot}}}=\frac{\mathcal{N}(x;\mu_x,\sigma_x^2)}{2} \left[1 + C \cos\big(\Delta\varphi + \kappa x\big)\right],
    \label{eq:pr-model}
\end{equation}
where \(\mathcal{N}(x;\mu_x,\sigma_x^2)\) is the Gaussian envelope and \(\kappa\) denotes the phase-shift gradient.
In general, the interferometric phase $\Delta\varphi(x,\bm{\theta})$ depends on the shot-to-shot parameters $\bm{\theta}$ through the spatially varying perturbation $\overline{\delta\varphi}(x,\bm{\theta})$. 
In contrast, the fitting model in Eq.~\eqref{eq:pr-model} treats $\Delta\varphi$ and $\kappa$ as free parameters that do not explicitly depend on $\bm{\theta}$. 
Consequently, the fitted values $\hat{\Delta\varphi}$ and $\hat{\kappa}$ correspond to biased estimates of the ideal parameters $\Delta\varphi_0$ and $\kappa_0$, 
defined in the absence of wavefront aberrations. 
The magnitude of this bias is determined by the unobserved variations in $\bm{\theta}$.

Expanding Eq.~\eqref{eq:perturbation-betas} linearly about $x=\mu_x$, one can see that 
the phase shift and phase shift gradient will be systematically biased:
\begin{equation}
\begin{gathered}
    \Delta\varphi = \Delta\varphi_{0}+\beta_{0}(\bm{\theta}) - \beta_{2}(\bm{\theta})\mu_{x}^2, \\
    \kappa = \kappa_{0}+\beta_{1}(\bm{\theta}) + 2\beta_{2}(\bm{\theta})\mu_{x},\label{eq:linearized-params}
\end{gathered}
\end{equation}
where $\kappa_{0}$ is the applied phase gradient. In what follows, we denote by $\hat{\kappa}$ and $\hat{\mu}_x$ 
the single-shot estimators of the phase-shift gradient and COM position, with corresponding variances 
$\sigma^2_{\hat{\kappa}}=\mathrm{Var}[\hat{\kappa}]$ and $\sigma^2_{\hat{\mu}_x}=\mathrm{Var}[\hat{\mu}_x]$, respectively.

Figures~\ref{fig:phase-shear-fits} and~\ref{fig:phase-shear-phi-kappa} show results from 25 MC simulations of phase-shear readout 
measurements 
in a laser beam with a uniform intensity profile, to enforce
Gaussianity at detection time, but keeping the Gaussian wavefront curvature given by Eq.~\eqref{eq:gaussian-phase}. Note that we set $f=0$ to maximize the effect.
The extracted values
$\hat{\Delta\varphi}$ and $\hat{\kappa}$ are in good agreement with the analytical predictions.

Now, because $\beta_{1}$ is linear in both $\mu_x$ and $\mu_{v_x}$,
and $\mu_x$ can directly be extracted from the images,
measuring $\kappa$ indirectly gives access to $\mu_{v_x}$,
opening the possibility of correcting for wavefront-induced phase shifts
using only resolved quantities (in a position-averaged measurement the term linear in $\mu_{v_x}$
vanishes upon averaging, so this information is lost).
We tested this by extracting $\mu_x$, $\kappa$, and $\Delta\varphi$ from 
each simulation.
As shown in Fig.~\ref{fig:fitted-phase-perturbation}, the phase shift
varies smoothly with $(\kappa, \mu_x)$, confirming that wavefront-induced 
effects are learnable from the data. The root mean square (RMS) error of the
fit is $1.84$ mrad, consistent with the RMS bootstrap standard deviation 
of the extracted $\hat{\Delta\varphi}$ due to atom shot noise ($1.57$ mrad).
This indicates that the residuals are dominated by atom shot noise, 
and that $\kappa$ and $\mu_x$ are good predictors of the wavefront-induced phase shifts.

Interestingly, this also implies that, if variations in $\mu_{x_0}$ and $\mu_{v_{x0}}$ are experimentally controlled and $\Delta\varphi_0$ remains constant shot-to-shot, observed variations in $\Delta\varphi$ and $\kappa$ can be used as regression parameters (together with an appropriate model, such as the simulations presented here) to estimate the underlying beam parameters. Complementing existing in-situ characterisation techniques, 
such as probing the spatial distribution of wavevectors using Bose-Einstein 
condensates~\cite{gaudout2025probingspatialdistributionkvectors}, this 
approach would provide an alternative pathway toward systematic error 
estimation in precision measurements, for instance in determinations of 
the fine-structure constant~\cite{Bouchendira2010,Parker,Morel2020}.

Note that, while our derivation assumed Gaussianity at detection time, enforced in the simulations via a flat intensity profile, 
preliminary simulations using realistic (non-uniform) profiles suggest that the 
extracted phase shift is still a smooth function of $(\kappa,\mu_x)$, and 
so position-resolved readout could remain a viable strategy. But further study is needed to 
fully characterize the effect of non-uniform intensity profiles on the extracted phase shift.
Additionally, we note that if velocity spread $\sigma_v$ varies significantly between shots, the curvature 
term  $\beta_2$ may become relevant. While our model predicts a smooth dependence of $\beta_2$ on $\sigma_v$, we found it difficult to extract
due to strong correlations with $\beta_0$ and $\beta_1$. Partial measurements of the atom cloud (for example from 
atoms not participating in the interferometer sequence) could be used to obtain additional information 
on COM coordinates and spreads, helping to address these issues.

\subsection{Potential limitations}
\subsubsection{Measurement precision requirements}
The reported mitigation strategy relies on accurately resolving the parameters $\mu_x$ and $\kappa$ from position-resolved data.  
It can therefore reduce phase noise whenever the measurement precision in the above parameters exceeds the uncontrolled shot-to-shot fluctuations of the initial COM coordinates $\mu_{x_0}$ and $\mu_{v_{x0}}$.  
In this regime, measurement resolution, rather than initial-cloud control accuracy, becomes the dominant factor determining performance.  
The practical feasibility of the method will therefore depend on atom number, imaging resolution and stability, and data-analysis techniques, all of which can be optimized in specific experimental settings.

In our simulations, the mean bootstrap uncertainty in $\hat{\mu}_x$
was $0.8$ $\mu$m, already within the 
micron-level precision required to operate at the $10^{-5}$ rad 
level. Thus, the resolution in $\hat{\kappa}$ will likely be the 
main constraint. 
The sensitivity of $\kappa$
to fluctuations in $\mu_{v_{x0}}$ is given by
$\frac{\partial\kappa}{\partial\mu_{v_{x0}}}$, requiring a $\kappa$-measurement precision
\begin{equation}
    \sigma_{\hat{\kappa}}<\bigg|\frac{\partial\kappa}{\partial\mu_{v_{x0}}}\bigg|\Delta{\mu_{v_{x0}}}.\label{eq:condition}
\end{equation}
In our simulations we find 
$\big|\frac{\partial\kappa}{\partial\mu_{v_{x0}}}\big|= 2\times 10^{5}$ rad m$^{-2}$s
and $\sigma_{\hat{\kappa}}=0.72$ mrad/mm (average across all 
simulations), allowing for mitigation of fluctuations as small as 
$\Delta\mu_{v_{x0}}\sim4$ $\mu$ms$^{-1}$.
With $\sigma_{\hat{\kappa}}\propto N^{-1/2}$ and $\big|\frac{\partial\kappa}{\partial\mu_{v_{x0}}}\big|\sim n$,
we expect that scaling the experiment up to $N=10^9$, $n=1001$, would enable 
mitigation of fluctuations as small as $0.04$ $\mu$ms$^{-1}$, well within the 
$10^{-5}$ rad phase noise parameter space. In fact, given the scaling of 
$\sigma_{\hat{\kappa}}$ with $N$, and assuming that the COM velocity 
control requirements are set by atom shot noise, so $\Delta\mu_{v_{x0}}\sim \sigma_{\text{asn}}^{1/2}\propto N^{-1/4}$,
the condition in Eq.~\eqref{eq:condition} is always satisfied provided $N$ is large enough.
Also, note that the approximate scaling $\big|\frac{\partial \kappa}{\partial \mu_{v_{x0}}}\big|\sim \frac{n}{w_0^4}\big(\frac{\sigma_v}{\sigma_x}\big)^2$
(with other parameters fixed) suggests that while increasing $w_0$ reduces sensitivity, 
the response can be optimized by adjusting the ratio $\sigma_v / \sigma_x$ 
or by increasing $n$, within realistic experimental limits.  
These adjustments offer practical handles for improving sensitivity rather than strict constraints.  
Nonetheless, position-selection effects arising from non-uniform beam intensity 
may reduce $\kappa$'s sensitivity to $\mu_{v_{x0}}$ unless high-fidelity LMT pulses are employed (e.g., using composite pulses~\cite{Genov2014}). 
Characterizing this effect will be an important subject for future work.

Finally, although our analysis focuses on transverse motion along $x$, where the phase gradient $\kappa$ is intentionally large for phase-shear readout, 
the approach can naturally be extended to two dimensions.  
In this case, the phase gradient becomes a vector $\boldsymbol{\kappa} = (\kappa_x, \kappa_y)$, 
and both components can be extracted by appropriately imaging the atom cloud in the transverse plane, 
for instance by using cameras positioned at multiple viewing angles or from below. 
This would enable simultaneous sensitivity to motion along both $x$ and $y$, 
without requiring modifications to the interferometer geometry or laser alignment.

\subsubsection{Robustness to other couplings}
The present analysis assumes phase-gradient fluctuations arise solely from 
coupling between transverse motion and curved wavefronts. In practice, 
other mechanisms may contribute.
For example, the Coriolis force can imprint a phase gradient across the atom cloud~\cite{PhysRevLett.111.083001}. 
While future long-baseline interferometers are expected to employ 
Coriolis compensation schemes, residual compensation errors 
could still introduce phase-gradient noise.

Finally, we note that even in the general case when $\overline{\delta\varphi}$
is no longer fully determined by some resolved parameters 
$\bm{\theta}_{\text{res}}$ due to some environmental or technical noise, we can still reduce 
correlated noise by constructing a corrected phase shift estimate for the $i^{\text{th}}$ experimental shot:
\begin{equation}
    \hat{\Delta\varphi}_{\text{corr},i} = \hat{\Delta\varphi}_i - \mathbb{E}[\hat{\Delta\varphi}|\bm{\theta}_{\text{res},i}],
\end{equation}
where \(\mathbb{E}[\cdot|\bm{\theta}_{\text{res},i}]\) denotes the conditional expectation with respect to the resolved parameters for that shot.
If $\bm{\theta}_{\text{res}}$ is uncorrelated with the science signal, this correction reduces noise without biasing the signal, 
as long as $\mathbb{E}[\hat{\Delta\varphi}|\bm{\theta}_{\text{res},i}]$ can be estimated from the data.

\section{Conclusion}

In this work, we investigated the phase noise arising from shot-to-shot fluctuations in the atoms' initial transverse motion in the presence of the wavefront curvature in Gaussian laser beams. Starting from a semi-classical framework, we derived analytical expressions 
for the effective phase perturbation in position-averaged measurements, quantifying how 
initial COM position and velocity fluctuations bias the interferometric signal in long-baseline, large momentum transfer atom interferometers. 
These predictions were validated using Monte Carlo simulations, showing good 
agreement.

We demonstrated that curvature-induced phase noise depends strongly on the beam’s focus position $f$. Moving the laser focus from $z=0$ to $|z|=z_R$ can reduce curvature noise by approximately two orders of magnitude by suppressing longitudinal variations of the wavefront curvature. However, in atom gradiometers, this improvement comes at the cost of reduced Rabi frequency uniformity and degraded gradiometer $\pi$-pulse efficiency. In particular, realizing LMT sequences with $n\sim 1000$ is not feasible in $|f|=z_R$ configurations with baseline $L=100$ m and beam waists smaller than 5 cm unless independent pulse-area error mitigation techniques are implemented, irrespective of the available laser power. We quantified this trade-off and introduced two limiting regimes: the low-efficiency/low-noise (LELN, $f=z_R$) and high-efficiency/high-noise (HEHN, $f=0$) regimes. Future long-baseline detectors will likely operate between these limits, balancing curvature suppression with practical power and efficiency constraints.

We applied our results to estimate phase noise in 100-m and 1-km gradiometers modelled after upcoming long-baseline
atom interferometry experiments such as AION~\cite{Badurina_2020} and MAGIS-100~\cite{Abe_2021}.
We found that to achieve sub-$10^{-5}$~rad phase noise levels (the target sensitivity for these experiments) for a 100-m detector with $w_0=1$~cm, stability at the micron and sub-micron-per-second levels is necessary in the HEHN configuration, while tens-of-microns and few-microns-per-second stability suffice for the LELN case. This imposes stringent requirements on source preparation
stability in future long-baseline atom interferometry experiments, and could
necessitate significant improvements over current performances. 

As a potential solution, we
proposed a mitigation strategy based on position-resolved phase-shift 
readout. We demonstrated, both analytically and via MC simulations, that the wavefront-induced phase bias exhibits 
a smooth, deterministic dependence on the atom cloud's final COM position $\mu_x$ and 
the observed phase-shift gradient $\kappa$. This enables empirical learning and 
correction of wavefront-induced phase shifts without prior knowledge of the beam or atom source 
details. We showed that the accuracy with which $\kappa$ and $\mu_x$ can be measured 
becomes the limiting factor for noise mitigation, but preliminary analytical and 
simulation results indicate that the required measurement precision is attainable.
We also discussed the impact of additional effects such as non-uniform 
intensity profiles, fluctuating velocity spreads, and other sources of 
phase-shift gradient fluctuations. These considerations motivate further work to characterize 
such effects and to develop robust inference methods suitable for practical 
implementation.

Overall, this study provides a quantitative framework for 
assessing phase noise in atom interferometers arising from 
transverse atomic motion in curved wavefronts. Our results 
have direct implications for the design and operation of next-generation, 
long-baseline, time-resolved atom interferometers for fundamental physics 
applications, including gravitational wave detection and ultra-light dark matter 
searches. Furthermore, the mitigation strategy we propose could play a key role 
in enabling these experiments to reach their target sensitivities, thereby 
advancing the frontiers of precision quantum sensing in fundamental physics.

\section*{Acknowledgments}
This work was supported by UKRI STFC under grant numbers ST/T006579/1, ST/W006200/1, and ST/X004864/1, as well as by the Isaac Newton Trust.  
We thank members of the AION and MAGIS-100 collaborations for their valuable input, and in particular Tim Kovachy, Michael Kagan, and Jonathan Tinsley for insightful discussions and feedback on this work.

\bibliographystyle{unsrt}
\bibliography{bibclean}

@inproceedings{hogan2008lightpulseatominterferometry,
  title         = {{Light-pulse atom interferometry}},
  author        = {Hogan, J. M. and David M. S. Johnson and Kasevich, M. A.},
  booktitle     = {Proceedings of the International Summer School of Physics ``Enrico Fermi'': Atom Optics and Space Physics},
  year          = {2008},
  note          = {To appear in the Proceedings of the International Summer School of Physics ``Enrico Fermi'' (Varenna, July 2007)},
  eprint        = {0806.3261},
  archivePrefix = {arXiv},
  primaryClass  = {physics.atom-ph},
  url           = {https://arxiv.org/abs/0806.3261}
}

@article{Sugarbaker_2013,
   title={{Enhanced Atom Interferometer Readout through the Application of Phase Shear}},
   volume={111},
   ISSN={1079-7114},
   url={http://dx.doi.org/10.1103/PhysRevLett.111.113002},
   number={11},
   journal={Physical Review Letters},
   publisher={American Physical Society (APS)},
   author={Sugarbaker, A. and Dickerson, S. M. and Hogan, J. M. and Johnson, D. M. S. and Kasevich, M. A.},
   year={2013},
   month=sep }

@article{Pagot2024,
author = {Pagot, L. and Merlet, S. and Dos Santos, F. P.},
journal = {Opt. Express},
number = {9},
pages = {18843--18854},
publisher = {Optica Publishing Group},
title = {{Influence of optical aberrations on the accuracy of an atomic gravimeter}},
volume = {33},
month = {May},
year = {2025},
url = {https://opg.optica.org/oe/abstract.cfm?URI=oe-33-9-18843},
}

@phdthesis{antoine2004,
  author       = {Antoine, C.},
  title        = {{Contribution à la théorie des interféromètres atomiques}},
  school       = {Université Pierre et Marie Curie - Paris VI},
  year         = {2004},
  url          = {https://tel.archives-ouvertes.fr/tel-00007967v2}
}

@article{antoine2006,
  author       = {Antoine, C.},
  title        = {{Matter wave beam splitters in gravito-inertial and trapping potentials: generalized ttt scheme for atom interferometry}},
  journal      = {Applied Physics B},
  volume       = {84},
  pages        = {585--597},
  year         = {2006},
}

@article{Lan2012,
   author = {Lan, S. Y. and Kuan, P. C. and Estey, B. and Haslinger, P. and Müller, H.},
   issn = {00319007},
   issue = {9},
   journal = {Physical Review Letters},
   month = {2},
   title = {{Influence of the coriolis force in atom interferometry}},
   volume = {108},
   year = {2012}
}

@Inbook{Svelto2010,
author="Svelto, O.",
title="Ray and Wave Propagation Through Optical Media",
bookTitle="Principles of Lasers",
year="2010",
publisher="Springer US",
address="Boston, MA",
pages="131--161",
isbn="978-1-4419-1302-9",
url="https://doi.org/10.1007/978-1-4419-1302-9_4"
}

@article{Abe_2021,
   title={{Matter-wave Atomic Gradiometer Interferometric Sensor (MAGIS-100)}},
   volume={6},
   ISSN={2058-9565},
   url={http://dx.doi.org/10.1088/2058-9565/abf719},
   number={4},
   journal={Quantum Science and Technology},
   publisher={IOP Publishing},
   author={Abe, M. and Adamson, P. and Borcean, M. and Bortoletto, D. and Bridges, K. and Carman, Samuel P et al.},
   year={2021},
   month=jul, pages={044003} }

@article{Badurina_2020,
   title={{AION: an atom interferometer observatory and network}},
   volume={2020},
   ISSN={1475-7516},
   url={http://dx.doi.org/10.1088/1475-7516/2020/05/011},
   number={05},
   journal={Journal of Cosmology and Astroparticle Physics},
   publisher={IOP Publishing},
   author={Badurina, L. and Bentine, E. and Blas, D. and Bongs, K. and Bortoletto, D. et al.},
   year={2020},
   month=may, pages={011–011} }

@article{
Parker,
author = {Parker, R. H. and Yu, C. and Zhong, W. and Estey, B. and Müller, H.},
title = {{Measurement of the fine-structure constant as a test of the Standard Model}},
journal = {Science},
volume = {360},
number = {6385},
pages = {191-195},
year = {2018},
URL = {https://www.science.org/doi/abs/10.1126/science.aap7706},
eprint = {https://www.science.org/doi/pdf/10.1126/science.aap7706},
}

@Inbook{MeystreTextBook,
author="Meystre, P. A. S. M.",
editor="Meystre, Pierre
and Sargent, Murray",
title="Mechanical Effects of Light",
bookTitle="Elements of Quantum Optics",
year="2007",
publisher="Springer Berlin Heidelberg",
address="Berlin, Heidelberg",
pages="151--170",
isbn="978-3-540-74211-1",
url="https://doi.org/10.1007/978-3-540-74211-1_6"
}

@article{Moore1997,
  title = {{Effects of atomic diffraction on the collective atomic recoil laser}},
  author = {Moore, M. G. and Meystre, P.},
  journal = {Phys. Rev. A},
  volume = {58},
  issue = {4},
  pages = {3248--3258},
  numpages = {0},
  year = {1998},
  month = {Oct},
  publisher = {American Physical Society},
  doi = {10.1103/PhysRevA.58.3248},
  url = {https://link.aps.org/doi/10.1103/PhysRevA.58.3248}
}

@article{Bade2018,
   author = {Bade, S. and Djadaojee, L. and Andia, M. and Cladé, P. and Guellati-Khelifa, S.},
   issn = {10797114},
   issue = {7},
   journal = {Physical Review Letters},
   month = {8},
   pmid = {30169104},
   publisher = {American Physical Society},
   title = {{Observation of Extra Photon Recoil in a Distorted Optical Field}},
   volume = {121},
   year = {2018}
}

@article{Storey1994,
   author = {Storey, P. and Cohen-Tannoudji, C.},
   issue = {11},
   journal = {Journal de Physique II},
   pages = {1999-2027},
   title = {{The Feynman path integral approach to atomic interferometry. A tutorial}},
   volume = {4},
   url = {https://hal.science/jpa-00248106},
   year = {1994}
}

@book{eaton2007multivariate,
  author    = {Eaton, M. L.},
  title     = {{Multivariate Statistics: A Vector Space Approach}},
  series    = {Lecture Notes-Monograph Series},
  volume    = {53},
  year      = {2007},
  publisher = {Institute of Mathematical Statistics},
  pages     = {i--512},
  url       = {http://www.jstor.org/stable/20461449},
}

@incollection{YOUNG1997363,
title = {{Precision Atom Interferometry with Light Pulses}},
editor = {Paul R. Berman},
booktitle = {Atom Interferometry},
publisher = {Academic Press},
address = {San Diego},
pages = {363-406},
year = {1997},
isbn = {978-0-12-092460-8},
url = {https://www.sciencedirect.com/science/article/pii/B9780120924608500102},
author = {Young, B. and Kasevich, M. and Chu, S.},
}

@article{Borde1989,
title = {{Atomic interferometry with internal state labelling}},
journal = {Physics Letters A},
volume = {140},
number = {1},
pages = {10-12},
year = {1989},
issn = {0375-9601},
doi = {https://doi.org/10.1016/0375-9601(89)90537-9},
url = {https://www.sciencedirect.com/science/article/pii/0375960189905379},
author = {Bordé, C.}
}

@article{Bouchendira2010,
  title = {{New Determination of the Fine Structure Constant and Test of the Quantum Electrodynamics}},
  author = {Bouchendira, R. and Clad\'e, P. and Guellati-Kh\'elifa, S. and Nez, F. M. C. and Biraben, F. M. C.},
  journal = {Phys. Rev. Lett.},
  volume = {106},
  issue = {8},
  pages = {080801},
  numpages = {4},
  year = {2011},
  month = {Feb},
  publisher = {American Physical Society},
  doi = {10.1103/PhysRevLett.106.080801},
  url = {https://link.aps.org/doi/10.1103/PhysRevLett.106.080801}
}

@article{Rosi_2014,
   title={{Precision measurement of the Newtonian gravitational constant using cold atoms}},
   volume={510},
   ISSN={1476-4687},
   url={http://dx.doi.org/10.1038/nature13433},
   number={7506},
   journal={Nature},
   publisher={Springer Science and Business Media LLC},
   author={Rosi, G. and Sorrentino, F. and Cacciapuoti, L. and Prevedelli, M. and Tino, G. M.},
   year={2014},
   month=jun, pages={518–521}
}

@article{peters1999,
  author    = {Peters, A. and Chung, K. Y. and Chu, S.},
  title     = {{Measurement of gravitational acceleration by dropping atoms}},
  journal   = {Nature},
  volume    = {400},
  pages     = {849--852},
  year      = {1999},
  url       = {https://doi.org/10.1038/23655}
}

@article{PhysRevA.88.043615,
  title = {{Simultaneous dual-species matter-wave accelerometer}},
  author = {Bonnin, A. and Zahzam, N. and Bidel, Y. and Bresson, A.},
  journal = {Phys. Rev. A},
  volume = {88},
  issue = {4},
  pages = {043615},
  numpages = {5},
  year = {2013},
  month = {Oct},
  publisher = {American Physical Society},
  url = {https://link.aps.org/doi/10.1103/PhysRevA.88.043615}
}

@article{PhysRevLett.113.023005,
  title = {{Test of Einstein Equivalence Principle for 0-Spin and Half-Integer-Spin Atoms: Search for Spin-Gravity Coupling Effects}},
  author = {Tarallo, M. G. and Mazzoni, T. and Poli, N. and Sutyrin, D. V. and Zhang, X. and Tino, G. M.},
  journal = {Phys. Rev. Lett.},
  volume = {113},
  issue = {2},
  pages = {023005},
  numpages = {5},
  year = {2014},
  month = {Jul},
  publisher = {American Physical Society},
  url = {https://link.aps.org/doi/10.1103/PhysRevLett.113.023005}
}

@article{PhysRevLett.78.2046,
  title = {{Precision Rotation Measurements with an Atom Interferometer Gyroscope}},
  author = {Gustavson, T. L. and Bouyer, P. and Kasevich, M. A.},
  journal = {Phys. Rev. Lett.},
  volume = {78},
  issue = {11},
  pages = {2046--2049},
  numpages = {0},
  year = {1997},
  month = {Mar},
  publisher = {American Physical Society},
  url = {https://link.aps.org/doi/10.1103/PhysRevLett.78.2046}
}

@article{PhysRevA.65.033608,
  title = {{Sensitive absolute-gravity gradiometry using atom interferometry}},
  author = {McGuirk, J. M. and Foster, G. T. and Fixler, J. B. and Snadden, M. J. and Kasevich, M. A.},
  journal = {Phys. Rev. A},
  volume = {65},
  issue = {3},
  pages = {033608},
  numpages = {14},
  year = {2002},
  month = {Feb},
  publisher = {American Physical Society},
  url = {https://link.aps.org/doi/10.1103/PhysRevA.65.033608}
}

@incollection{abend2020atominterferometryapplications,
  author       = {Abend, S. and Gersemann, M. and Schubert, C. and Schlippert, D. and Rasel, E. M. et al.},
  title        = {{Atom Interferometry and Its Applications}},
  booktitle    = {Foundations of Quantum Theory},
  series       = {Proceedings of the International School of Physics ``Enrico Fermi'', Course 197},
  pages        = {345--392},
  year         = {2019},
  publisher    = {IOS Press and Societ{\`a} Italiana di Fisica},
  address      = {Amsterdam; Bologna},
  doi          = {10.3254/ENFI240005},
}

@article{PhysRevLett.110.171102,
  title = {{New Method for Gravitational Wave Detection with Atomic Sensors}},
  author = {Graham, P. W. and Hogan, J. M. and Kasevich, M. A. and Rajendran, S.},
  journal = {Phys. Rev. Lett.},
  volume = {110},
  issue = {17},
  pages = {171102},
  numpages = {5},
  year = {2013},
  month = {Apr},
  publisher = {American Physical Society},
  url = {https://link.aps.org/doi/10.1103/PhysRevLett.110.171102}
}

@article{Hogan_2011,
   title={{An atomic gravitational wave interferometric sensor in low earth orbit (AGIS-LEO)}},
   volume={43},
   ISSN={1572-9532},
   url={http://dx.doi.org/10.1007/s10714-011-1182-x},
   number={7},
   journal={General Relativity and Gravitation},
   publisher={Springer Science and Business Media LLC},
   author={Hogan, J. M. and Johnson, D. M. S. and Dickerson, S. and Kovachy, T. and Sugarbaker, Alex et al.},
   year={2011},
   month=may, pages={1953–2009} 
}

@article{PhysRevLett.115.011802,
  title = {{Search for Ultralight Scalar Dark Matter with Atomic Spectroscopy}},
  author = {Van Tilburg, K. and Leefer, N. and Bougas, L. and Budker, D.},
  journal = {Phys. Rev. Lett.},
  volume = {115},
  issue = {1},
  pages = {011802},
  numpages = {5},
  year = {2015},
  month = {Jun},
  publisher = {American Physical Society},
  url = {https://link.aps.org/doi/10.1103/PhysRevLett.115.011802}
}

@article{PhysRevLett.117.261301,
  title = {{Sensitivity of Atom Interferometry to Ultralight Scalar Field Dark Matter}},
  author = {Geraci, A. A. and Derevianko, A.},
  journal = {Phys. Rev. Lett.},
  volume = {117},
  issue = {26},
  pages = {261301},
  numpages = {5},
  year = {2016},
  month = {Dec},
  publisher = {American Physical Society},
  url = {https://link.aps.org/doi/10.1103/PhysRevLett.117.261301}
}

@article{Badurina_2021,
   title={{Prospective sensitivities of atom interferometers to gravitational waves and ultralight dark matter}},
   volume={380},
   ISSN={1471-2962},
   url={http://dx.doi.org/10.1098/rsta.2021.0060},
   number={2216},
   journal={Philosophical Transactions of the Royal Society A: Mathematical, Physical and Engineering Sciences},
   publisher={The Royal Society},
   author={Badurina, L. and Buchmueller, O. and Ellis, J. and Lewicki, M. and McCabe, C. and Vaskonen, V.},
   year={2021},
   month=dec }

@article{Wicht2005,
   author = {Wicht, A. and Sarajlic, E. and Hensley, J. M. and Chu, S.},
   issn = {10502947},
   issue = {2},
   journal = {Physical Review A - Atomic, Molecular, and Optical Physics},
   month = {8},
   title = {{Phase shifts in precision atom interferometry due to the localization of atoms and optical fields}},
   volume = {72},
   year = {2005}
}

@article{Fils2005,
   author = {Fils, J. and Leduc, F. and Bouyer, P. and Holleville, D. and Dimarcq, N. and Clairon, A. and Landragin, A.},
   issn = {14346079},
   issue = {3},
   journal = {European Physical Journal D},
   pages = {257-260},
   publisher = {Springer New York},
   title = {{Influence of optical aberrations in an atomic gyroscope}},
   volume = {36},
   year = {2005}
}

@article{Karcher2018,
  title = {{Improving the accuracy of atom interferometers with ultracold sources}},
  author = {Karcher, R. and Aimanaliyev, A. and Merlet, S. and Pereira Dos Santos, F.},
  journal = {New Journal of Physics},
  volume = {20},
  number = {11},
  pages = {113041},
  year = {2018},
  publisher = {IOP Publishing},
  doi = {10.1088/1367-2630/aaf07d},
  url = {https://iopscience.iop.org/article/10.1088/1367-2630/aaf07d},
}

@article{Louchet-Chauvet2011,
   author = {Louchet-Chauvet, A. and Farah, T. and Bodart, Q. and Clairon, A. and Arnaud Landragin et al.},
   issue = {6},
   journal = {New Journal of Physics},
   month = {6},
   pages = {065025},
   publisher = {IOP Publishing},
   title = {{The influence of transverse motion within an atomic gravimeter}},
   volume = {13},
   year = {2011}
}

@inproceedings{canuel2022gravityantennabasedquantum,
  title        = {{A gravity antenna based on quantum technologies: MIGA}},
  author       = {Canuel, B. and Zou, X. and Sabulsky, D. O. and Junca, J. and Bertoldi, A. and others},
  booktitle    = {Proceedings of the 56th Rencontres de Moriond: Gravitation},
  year         = {2022},
  note         = {Contribution to the Gravitation session},
  eprint       = {2204.12137},
  archivePrefix= {arXiv},
  primaryClass = {physics.atom-ph},
  url          = {https://arxiv.org/abs/2204.12137}
}

@article{Zhao_2022,
   title={{Ultralight scalar dark matter detection with ZAIGA}},
   volume={31},
   ISSN={1793-6594},
   url={http://dx.doi.org/10.1142/S0218271822500377},
   number={05},
   journal={International Journal of Modern Physics D},
   publisher={World Scientific Pub Co Pte Ltd},
   author={Zhao, W. and Mei, X. and Gao, D. and Wang, J. and Zhan, M.},
   year={2022},
   month=feb }

@article{Miffre2006,
  title        = {{Atom interferometry}},
  author       = {Miffre, A. and Jacquey, M. and B{\"u}chner, M. and Tr{\'e}nec, G. and Vigu{\'e}, J.},
  journal      = {Physica Scripta},
  volume       = {74},
  number       = {2},
  pages        = {C15},
  year         = {2006},
  publisher    = {The Royal Swedish Academy of Sciences},
  doi          = {10.1088/0031-8949/74/2/N01},
  url          = {https://iopscience.iop.org/article/10.1088/0031-8949/74/2/N01}
}

@article{PhysRevA.74.023615,
  title = {{Atom interferometer as a selective sensor of rotation or gravity}},
  author = {Dubetsky, B. and Kasevich, M. A.},
  journal = {Phys. Rev. A},
  volume = {74},
  issue = {2},
  pages = {023615},
  numpages = {17},
  year = {2006},
  month = {Aug},
  publisher = {American Physical Society},
  url = {https://link.aps.org/doi/10.1103/PhysRevA.74.023615}
}

@inbook{quantumopticsphasespace,

publisher = {John Wiley and Sons},
isbn = {9783527602971},
title = {{Wigner Function}},
booktitle = {{Quantum Optics in Phase Space}},
chapter = {3},
pages = {67-98},
url = {https://onlinelibrary.wiley.com/doi/abs/10.1002/3527602976.ch3},
eprint = {https://onlinelibrary.wiley.com/doi/pdf/10.1002/3527602976.ch3},
year = {2001},
author = {Schleich, W. P.}
}

@article{Schkolnik_2015,
   title={{The effect of wavefront aberrations in atom interferometry}},
   volume={120},
   ISSN={1432-0649},
   url={http://dx.doi.org/10.1007/s00340-015-6138-5},
   number={2},
   journal={Applied Physics B},
   publisher={Springer Science and Business Media LLC},
   author={Schkolnik, V. and Leykauf, B. and Hauth, M. and Freier, C. and Peters, A.},
   year={2015},
   month=jun, pages={311–316} }

@misc{utz2014extendedwignerfunctionformalism,
      title={{Extended Wigner function formalism for the spatial propagation of particles with internal degrees of freedom}}, 
      author={Utz, M. and Levitt, M. H. and Cooper, N. and Ulbricht, H.},
      year={2014},
      eprint={1405.1886},
      archivePrefix={arXiv},
      primaryClass={quant-ph},
      url={https://arxiv.org/abs/1405.1886}, 
}

@article{PhysRevLett.67.181,
  title = {{Atomic interferometry using stimulated Raman transitions}},
  author = {Kasevich, M. and Chu, S.},
  journal = {Phys. Rev. Lett.},
  volume = {67},
  issue = {2},
  pages = {181--184},
  numpages = {0},
  year = {1991},
  month = {Jul},
  publisher = {American Physical Society},
  url = {https://link.aps.org/doi/10.1103/PhysRevLett.67.181}
}

@incollection{Tolman1938,
  author    = {Tolman, R. C.},
  title     = {{Liouville's Theorem for the Change in Density with Time}},
  booktitle = {The Principles of Statistical Mechanics},
  publisher = {Oxford University Press, Clarendon Press},
  address   = {Oxford},
  year      = {1938},
  chapter   = {III.19},
  pages     = {48--52}
}

@article{PhysRevLett.131.033602,
  title = {{Large-Momentum-Transfer Atom Interferometers with $\mathrm{\ensuremath{\mu}}\mathrm{rad}$-Accuracy Using Bragg Diffraction}},
  author = {Kirsten-Siem\ss{}, J.-N. and Fitzek, F. and Schubert, C. and Rasel, E. M. and Gaaloul, N. and et al.},
  journal = {Phys. Rev. Lett.},
  volume = {131},
  issue = {3},
  pages = {033602},
  numpages = {7},
  year = {2023},
  month = {Jul},
  publisher = {American Physical Society},
  url = {https://link.aps.org/doi/10.1103/PhysRevLett.131.033602}
}

@misc{baynham2025prototypeatominterferometerdetect,
  title         = {{A Prototype Atom Interferometer to Detect Dark Matter and Gravitational Waves}},
  author        = {C. F. A. Baynham and Hobson, R. and Buchmueller, O. and Evans, D. and L. Hawkins et al.},
  year          = {2025},
  eprint        = {2504.09158},
  archivePrefix = {arXiv},
  primaryClass  = {hep-ex},
  note          = {Preprint (arXiv:2504.09158)},
  url           = {https://arxiv.org/abs/2504.09158}
}

@article{seckmeyer2024principalcomponentanalysisspatial,
  author       = {Seckmeyer, S. and Ahlers, H. and Kirsten-Siemß, J.-N. and Gersemann, M. and Ernst M. Rasel et al.},
  title        = {{Spatially resolved phase reconstruction for atom interferometry}},
  journal      = {EPJ Quantum Technology},
  volume       = {12},
  number       = {1},
  pages        = {34},
  year         = {2025},
  doi          = {10.1140/epjqt/s40507-025-00337-2},
  url          = {https://doi.org/10.1140/epjqt/s40507-025-00337-2},
  issn         = {2196-0763}
}

@Inbook{Gut1995,
author="Gut, A.",
title="The Multivariate Normal Distribution",
bookTitle="An Intermediate Course in Probability",
year="1995",
publisher="Springer New York",
address="New York, NY",
pages="119--148",
isbn="978-1-4757-2431-8",
url="https://doi.org/10.1007/978-1-4757-2431-8_6"
}

@article{canuel2020technologieselgarlargescale,
  title        = {{ELGAR—a European Laboratory for Gravitation and Atom-interferometric Research}},
  author       = {Canuel, B. and Abend, S. and Amaro-Seoane, P. and Badaracco, F. and Q. Beaufils et al.},
  journal      = {Classical and Quantum Gravity},
  volume       = {37},
  number       = {22},
  pages        = {225017},
  year         = {2020},
  publisher    = {IOP Publishing},
  doi          = {10.1088/1361-6382/aba80e},
  url          = {https://iopscience.iop.org/article/10.1088/1361-6382/aba80e}
}

@article{Asenbaum_2020,
   title={{Atom-Interferometric Test of the Equivalence Principle at the 1e-12 Level}},
   volume={125},
   ISSN={1079-7114},
   url={http://dx.doi.org/10.1103/PhysRevLett.125.191101},
   number={19},
   journal={Physical Review Letters},
   publisher={American Physical Society (APS)},
   author={Asenbaum, P. and Overstreet, C. and Kim, M. and Curti, J. and Kasevich, M. A.},
   year={2020},
   month=nov }

@article{PhysRevA.101.033606,
  title = {{Accurate trajectory alignment in cold-atom interferometers with separated laser beams}},
  author = {Altorio, M. and Sidorenkov, L. A. and Gautier, R. and Savoie, D. and Landragin, A. et al.},
  journal = {Phys. Rev. A},
  volume = {101},
  issue = {3},
  pages = {033606},
  numpages = {10},
  year = {2020},
  month = {Mar},
  publisher = {American Physical Society},
  url = {https://link.aps.org/doi/10.1103/PhysRevA.101.033606}
}

@article{Dean2018,
   author = {Dean, D. S. and Le Doussal, P. and Majumdar, S. N. and Schehr, G.},
   issn = {24699934},
   issue = {6},
   journal = {Physical Review A},
   month = {6},
   publisher = {American Physical Society},
   title = {{Wigner function of noninteracting trapped fermions}},
   volume = {97},
   year = {2018}
}

@article{glick2023coriolisforcecompensationlaser,
  title        = {{Coriolis force compensation and laser beam delivery for 100-m baseline atom interferometry}},
  author       = {Glick, J. and Chen, Z. and Deshpande, T. and Wang, Y. and Kovachy, T.},
  journal      = {AVS Quantum Science},
  volume       = {6},
  number       = {1},
  pages        = {014402},
  year         = {2024},
  publisher    = {AIP Publishing},
  doi          = {10.1116/5.0180083},
  url          = {https://doi.org/10.1116/5.0180083}
}

@article{PhysRevLett.111.083001,
  title = {{Multiaxis Inertial Sensing with Long-Time Point Source Atom Interferometry}},
  author = {Dickerson, S. M. and Hogan, J. M. and Sugarbaker, A. and Johnson, D. M. S. and Kasevich, M. A.},
  journal = {Phys. Rev. Lett.},
  volume = {111},
  issue = {8},
  pages = {083001},
  numpages = {5},
  year = {2013},
  month = {Aug},
  publisher = {American Physical Society},
  url = {https://link.aps.org/doi/10.1103/PhysRevLett.111.083001}
}

@article{Chen2020,
author = {Chen, Y.-J. and Hansen, A. and Shuker, M. and Boudot, R. and John Kitching et al.},
journal = {Opt. Express},
number = {23},
pages = {34516--34529},
publisher = {Optica Publishing Group},
title = {{Robust inertial sensing with point-source atom interferometry for interferograms spanning a partial period}},
volume = {28},
month = {Nov},
year = {2020},
url = {https://opg.optica.org/oe/abstract.cfm?URI=oe-28-23-34516},
doi = {10.1364/OE.399988},
}

@article{PhysRevA.102.013326,
  title = {{Rotation sensing with improved stability using point-source atom interferometry}},
  author = {Avinadav, C. and Yankelev, D. and Shuker, M. and Firstenberg, O. and Davidson, N.},
  journal = {Phys. Rev. A},
  volume = {102},
  issue = {1},
  pages = {013326},
  numpages = {8},
  year = {2020},
  month = {Jul},
  publisher = {American Physical Society},
  url = {https://link.aps.org/doi/10.1103/PhysRevA.102.013326}
}

@article{Genov2014,
   author = {Genov, G. T. and Schraft, D. and Halfmann, T. and Vitanov, N. V.},
   doi = {10.1103/PhysRevLett.113.043001},
   issn = {10797114},
   issue = {4},
   journal = {Physical Review Letters},
   month = {7},
   publisher = {American Physical Society},
   title = {{Correction of arbitrary field errors in population inversion of quantum systems by universal composite pulses}},
   volume = {113},
   year = {2014}
}

@article{Morel2020,
  author    = {Morel, L. and Yao, Z. and Cladé, P. and Guellati-Khélifa, S.},
  title     = {{Determination of the fine-structure constant with an accuracy of 81 parts per trillion}},
  journal   = {Nature},
  year      = {2020},
  volume    = {588},
  number    = {7836},
  pages     = {61--65},
  doi       = {10.1038/s41586-020-2964-7},
  url       = {https://doi.org/10.1038/s41586-020-2964-7}
}

@article{Buchmueller03042023,
author = {Buchmueller, O. and Ellis, J. and Schneider, U.},
title = {{Large-scale atom interferometry for fundamental physics}},
journal = {Contemporary Physics},
volume = {64},
number = {2},
pages = {93--110},
year = {2023},
publisher = {Taylor \& Francis},
doi = {10.1080/00107514.2023.2239008},


URL = { 
    
        https://doi.org/10.1080/00107514.2023.2239008
    
    

},
eprint = { 
    
        https://doi.org/10.1080/00107514.2023.2239008
    
    

}

}

@misc{gaudout2025probingspatialdistributionkvectors,
      title={{Probing the spatial distribution of k-vectors in situ with Bose-Einstein condensates}}, 
      author={Gaudout, S. and Si-Ahmed, R. and Debavelaere, C. and Door, M. and Cladé, P. and Guellati-Khelifa, S.},
      year={2025},
      eprint={2507.19157},
      archivePrefix={arXiv},
      primaryClass={physics.atom-ph},
      url={https://arxiv.org/abs/2507.19157}, 
}

\appendix

\section{Effect of longitudinal motion}\label{app:long-motion}
We here show that the effect of longitudinal motion is negligible in comparison to that of transverse motion.

In deriving Eq.~\eqref{eq:one-atom-phase}, we ignored the contribution from the Gouy phase. Using Eq.~\eqref{eq:lmt-phase-shift}, and setting $f=0$, we find that in a LMT MZ sequence the Gouy phase term in Eq.~\eqref{eq:gaussian-phase} contributes a phase shift \begin{equation}
\delta\varphi_{\text{Gouy}}(\bm{x}_0,\bm{v}_0) \approx -\frac{ngT^2}{z_R} - \frac{n}{3z_R^3}(z_0^3 - 2z_T^3 + z_{2T}^3),
\end{equation}
up to corrections of order $\mathcal{O}(1/z_R^5)$. We use $f=0$ as this corresponds to the regime where the sensitivity to longitudinal motion is approximately maximized (the true maximum lies close to $f=0$, and in any case when $f \ll z_R$). For the parameters of Fig.~\ref{fig:teff-plots}, and even with a conservative $z_0=1$ km, the sensitivities of this contribution evaluate to $\partial \delta\varphi{\text{Gouy}}/\partial z_0 \sim \mathcal{O}(10^{-3}$ mrad/mm) and $\partial \delta\varphi_{\text{Gouy}}/\partial v_{z_0} \sim \mathcal{O}(10^{-3}$ mrad/mm s$^{-1})$, which are three to four orders of magnitude smaller than the transverse sensitivities shown in Fig.~\ref{fig:teff-plots}.

The curvature term $k(x^2+y^2)/[2R(z)]$ also carries a weak $z$-dependence. For the same parameters, setting $f=0$, and assuming $x_0=1$ mm and $v_{x_0}=1$ mms$^{-1}$ (conservative estimate) we find that the sensitivity of $\delta\varphi$ (Eq.~\eqref{eq:one-atom-phase}) to fluctuations in the initial longitudinal coordinates evaluates to $\partial \delta\varphi/\partial z_0 \sim \mathcal{O}(10^{-3}$ mrad/mm) and $\partial \delta\varphi/\partial v_{z_0} \sim \mathcal{O}(10^{-2}$ mrad/mm s$^{-1})$, again far below the transverse sensitivities. 

We therefore treat the longitudinal coordinates as fixed from shot to shot and focus exclusively on the role of transverse motion.

\section{Monte Carlo simulation}\label{app:simulations}
To validate the analytical models derived in this work, we performed Monte Carlo (MC) simulations of a large momentum
transfer Mach-Zehnder atom interferometer.
Each simulation evaluates the total interferometric phase shift $\Delta\varphi$, given by equation~\eqref{eq:tot-phase-shift-one-wp}, for a 
large ensemble of individual trajectories with initial positions and velocities drawn from a Gaussian distribution,
and evolved by numerically solving Hamilton’s equations of motion. The propagation phase in equation~\eqref{eq:propagation-phase} is computed via numerical integration by evaluating the classical action along each trajectory. 
The quantum evolution of the internal states is modelled by solving the time-dependent Schrödinger equation 
for a Gaussian wave packet with two internal degrees of freedom in a plane wave with phase offset $\phi(\bm{x}_i)$, where $\bm{x}_i$ is the wave packet's central coordinate at the interaction time, following the approach in 
Refs.~\cite{antoine2004, antoine2006}. The simulations use a position-dependent Rabi frequency to represent 
arbitrary laser intensity profiles.
At the end of the interferometry sequence, trajectories that overlap in phase space and share the same internal state 
interfere, forming an effective \emph{interferometry port}. The probability of detecting an atom at a given port is:
\[
P = |A_1|^2 + |A_2|^2 + 2|A_1||A_2|\cos(\Delta\varphi),
\]
where \(A_1\) and \(A_2\) are the complex amplitudes associated with the interfering trajectories, and \(\Delta\varphi\) 
is their accumulated phase difference.

Each detected atom can be sampled from the resulting probability distribution across ports, yielding a dataset of \(N\) 
samples in phase and state space, denoted \({\{ \bm{x}_j, \bm{v}_j, s_j \}}_{j=1}^N\).
These samples are effectively used to approximate the distribution of the atoms as 
\begin{equation}
    p(s,\bm{x},\bm{v}) \approx \frac{1}{N}\sum_{j=1}^N \delta(\bm{x}-\bm{x}_j)\delta(\bm{v}-\bm{v}_j)\delta_{s s_j},
\end{equation}
which can then be used to the estimate the expected measurement outcome for arbitrary operators, enabling the validation of the analytical models derived in this work.


\section{Derivation of the effective phase perturbation in a position-resolved phase measurement for a Gaussian beam}\label{app:bias-gaussian-beam}

We assume that the positions and velocities of the atoms at $t=0$ are independently normally distributed,
\begin{equation}
    \begin{pmatrix}
        \bm{x}_0 \\ \bm{v}_0
    \end{pmatrix}
    \sim
    \mathcal{N}\bigg(
    \begin{pmatrix}
        \bm{\mu}_{x_0}\\
        \bm{\mu}_{v_0}
    \end{pmatrix}
    ,
    \begin{pmatrix}
        \sigma_{x_0}^2\mathbb{I}&0\\0&\sigma_{v_0}^2\mathbb{I}
    \end{pmatrix}
    \Bigg),
\end{equation}
where $\mathcal{N}(\bm{\mu},\bm{\Sigma})$ denotes a normal distribution with mean $\bm{\mu}$ and covariance matrix 
$\bm{\Sigma}$, and $\mathbb{I}$ is a $3\times 3$ unit matrix. Under the assumption that the atoms evolve in a uniform 
gravitational field with gravitational acceleration $\bm{g}$, classical equations of motion relate the initial phase-space 
coordinates to the detection-time phase-space coordinates by the linear transformation:
\begin{equation}
    \begin{pmatrix}
        \bm{x} \\ \bm{v}
    \end{pmatrix}=
    \begin{pmatrix}
        \mathbb{I} & t_{\text{det}}\\
        0 & \mathbb{I}
    \end{pmatrix}
    \cdot
    \begin{pmatrix}
        \bm{x}_0 \\ \bm{v}_0
    \end{pmatrix}
    +
    \begin{pmatrix}
        \frac{1}{2}\bm{g}t_{\text{det}}^2\\
        \bm{g}t_{\text{det}}
    \end{pmatrix},
\end{equation}
where $t_{\text{det}}$ is the detection time, and where we ignore the vertical separation between the interferometer's arms.
As a result of the linear relation, the phase-space coordinates at detection time are also normally distributed, 
and it is easy to show that
\begin{equation}
    \begin{pmatrix}
        \bm{x} \\\bm{v}
    \end{pmatrix}
    \sim
    \mathcal{N}\bigg(
    \begin{pmatrix}
        \bm{\mu}_{x}\\
        \bm{\mu}_{v}
    \end{pmatrix}
    ,
    \begin{pmatrix}
        \sigma_{x}^2\mathbb{I}&\sigma_{v}^2t_{\text{det}}\mathbb{I} \\\sigma_{v}^2t_{\text{det}}\mathbb{I}&\sigma_{v}^2\mathbb{I}
    \end{pmatrix}
    \bigg),
\end{equation}
with $\bm{\mu}_{x}=\bm{\mu}_{x_0}+\bm{\mu}_{v_0}t_{\text{det}}+\frac{1}{2}\bm{g}t_{\text{det}}^2$, $\bm{\mu}_{v}=\bm{\mu}_{v_0}+\bm{g}t_{\text{det}}$,
$\sigma_{x}^2=\sigma_{x_0}^2+\sigma_{v_0}^2t_{\text{det}}^2$ and $\sigma_{v}=\sigma_{v_0}$. 

In the one-dimensional position-resolved case, all phase-space coordinates but $x$ are integrated over. For the sake of 
conciseness, we denote the integrated coordinates $\bm{\xi}_I={(y,z,v_x,v_y,v_z)}^T$. Using properties of normal 
distributions (see Ref.~\cite{eaton2007multivariate}) it can then be shown that the integrated coordinates conditioned on $x$ are 
also normally distributed, i.e., $\bm{\xi}_I|x\sim \mathcal{N}(\bm{\mu}_{\xi _I|x},\bm{\Sigma}_{\xi _I|x})$, with
\begin{equation}
   \bm{\mu}_{\xi _I|x}=\begin{pmatrix}
       \mu_{y}\\
       \mu_{z}\\
       \mu_{v_{x}}+\frac{\sigma_{v}^2t_{\text{det}}}{\sigma_{x}^2}(x-\mu_{x})\\
       \mu_{v_y}\\
       \mu_{v_{z}}
   \end{pmatrix} 
\end{equation}
and

\begin{equation}
    \bm{\Sigma}_{\xi _I|x}=
    \begin{pmatrix}
        \sigma_{x}^2\mathbb{I} & \bm{0} & \sigma_{v}^2t_{\text{det}}\mathbb{I}\\
        \bm{0}^T & \sigma_{v}^2-\frac{\sigma_{v}^4t_{\text{det}}^2}{\sigma_{x}^2} & \bm{0}^T \\
        \sigma_{v}^2t_{\text{det}}\mathbb{I} & \bm{0} & \sigma_{v}^2\mathbb{I} \\
    \end{pmatrix},
\end{equation}
where $\mathbb{I}$ is here a $2\times2$ unit matrix and $\bm{0}={(0,0)}^T$.

Restricting our attention to the $x$-direction, we can express the one-atom perturbation $\delta\varphi$ in 
terms of final phase-space coordinates using Eq.~\eqref{eq:one-atom-phase}:
\begin{equation}
\begin{gathered}
    \delta\varphi(x,v_x)=\frac{1}{2}T^2\bigg(c_1^+(n+1)-c_1^-(n-1)\bigg)\\\times
    \bigg(C_{xx}x^2+C_{xv}xv_x+C_{vv}v_x^2\bigg),
\end{gathered}
\end{equation}
in which the coefficients $C_{ij}$ are defined as
\begin{align}
    C_{xx}&=C_{xx}^{(0)},\\
    C_{xv}&=C_{xv}^{(0)}-2C_{xx}^{(0)}t_{\text{det}},\\
    C_{vv}&=C_{vv}^{(0)}-C_{xv}^{(0)}t_{\text{det}}+C_{xx}^{(0)}t_{\text{det}}^2,
\end{align}
where the coefficients $C_{ij}^{(0)}$ are defined in Table~\ref{tab:phase-shift-coefficients} in the main text.

The expected value of the phase shift, conditioned on $x$, is then given by
\begin{equation}
\begin{gathered}
    \overline{\delta\varphi}(x,\bm{\theta})=\beta_0(\bm{\theta})+\beta_1(\bm{\theta}) x+\beta_2(\bm{\theta}) x^2,
\end{gathered}
\end{equation}
with
\begin{align}
    \beta_0(\mu_x,\mu_{v_x})&=C_0 C_{vv}(\mu_{v_x}-\mu_x\frac{\sigma_v^2}{\sigma_x^2}t_{\text{det}})^2,\\
    \beta_1(\mu_x,\mu_{v_x})&=C_0 (2C_{vv}\frac{\sigma_v^2}{\sigma_x^2}t_{\text{det}}+C_{xv})(\mu_{v_x}-\mu_x\frac{\sigma_v^2}{\sigma_x^4}t_{\text{det}}),\\
    \beta_2(\mu_x,\mu_{v_x})&=C_0 (C_{xx}+\frac{\sigma_v^2}{\sigma_x^2}t_{\text{det}}(C_{vv}\frac{\sigma_v^2}{\sigma_x^2}t_{\text{det}}+C_{xv})),
\end{align}
where we defined the common factor $C_0=\frac{1}{2}T^2\big(c_1^+(n+1)-c_1^-(n-1)\big)$.

\end{document}